\documentclass[superscriptaddress,amsmath,amssymb,aps,longbibliography,prb,preprint]{revtex4-2}
\usepackage{xr-hyper}  
\usepackage{hyperref} 

\usepackage{lineno}
\usepackage{graphicx}%
\usepackage{multirow}%
\usepackage{amsmath,amssymb,amsfonts}%
\usepackage{amsthm}%
\usepackage{mathrsfs}%
\usepackage[title]{appendix}%
\usepackage{xcolor}%
\usepackage{textcomp}%
\usepackage{booktabs}%
\usepackage{threeparttable}
\usepackage{algorithm}%
\usepackage{algorithmicx}%
\usepackage{algpseudocode}%
\usepackage{listings}%
\usepackage{orcidlink}%
\usepackage{array}
\usepackage{makecell}
\usepackage{tabularx}
\usepackage{bbding} 
\usepackage{longtable} 
\usepackage{colortbl}
\usepackage{pgfplots}  
\usepgfplotslibrary{colormaps}           
\pgfplotsset{compat=1.17,colormap/viridis}                 
\usepackage{pgfplotstable}         
\PassOptionsToPackage{table}{xcolor}
\usepackage[noabbrev,nameinlink,capitalise]{cleveref}
\usepackage[indent,parfill]{parskip}

\newcommand{\corres}[1]{}

\newcommand{\domainim}{\textit{Inorganic Materials}}
\newcommand{\domainsm}{\textit{Molecules}}
\newcommand{\domaincat}{\textit{Catalysis}}

\definecolor{viridis0}{RGB}{68,1,84}
\definecolor{viridis10}{RGB}{72,35,116}
\definecolor{viridis20}{RGB}{64,67,135}
\definecolor{viridis30}{RGB}{52,94,141}
\definecolor{viridis40}{RGB}{41,120,142}
\definecolor{viridis50}{RGB}{33,145,140}
\definecolor{viridis60}{RGB}{59,174,118}
\definecolor{viridis70}{RGB}{99,200,99}
\definecolor{viridis80}{RGB}{170,220,50}
\definecolor{viridis90}{RGB}{233,231,36}
\definecolor{viridis100}{RGB}{253,231,37}

\newcommand{\colormap}[1]{%
    \ifdim #1 pt < 0.1pt \color{black}\cellcolor{viridis100}%
    \else\ifdim #1 pt < 0.2pt \color{black}\cellcolor{viridis90}%
    \else\ifdim #1 pt < 0.3pt \color{black}\cellcolor{viridis80}%
    \else\ifdim #1 pt < 0.4pt \color{black}\cellcolor{viridis70}%
    \else\ifdim #1 pt < 0.5pt \color{black}\cellcolor{viridis60}%
    \else\ifdim #1 pt < 0.6pt \color{white}\cellcolor{viridis50}%
    \else\ifdim #1 pt < 0.7pt \color{white}\cellcolor{viridis40}%
    \else\ifdim #1 pt < 0.8pt \color{white}\cellcolor{viridis30}%
    \else\ifdim #1 pt < 0.9pt \color{white}\cellcolor{viridis20}%
    \else\ifdim #1 pt < 1.0pt \color{white}\cellcolor{viridis10}%
    \else \color{white}\cellcolor{viridis0}%
    \fi\fi\fi\fi\fi\fi\fi\fi\fi\fi%
}

\newcommand{\normcolor}[3]{%
  \pgfmathsetmacro{\normval}{(#1 - #2)/(#3 - #2)}%
  \edef\tempval{\normval}%
  \colormap{\tempval}#1%
}

\begin{document}

\title{LAMBench: A Benchmark for Large Atomistic Models}



\author{\firstname{Anyang} \surname{Peng}\corres{}\orcidlink{0000-0002-0630-2187}$^\dagger$}\email{pengay@aisi.ac.cn}
\affiliation{AI for Science Institute, Beijing, China}

\author{\firstname{Chun} \surname{Cai}\orcidlink{0000-0001-6242-0439}}\thanks{These authors contributed equally to this work.}

\affiliation{AI for Science Institute, Beijing, China}

\author{\firstname{Mingyu} \surname{Guo}\orcidlink{0009-0008-3744-1543}}
\affiliation{AI for Science Institute, Beijing, China}
\affiliation{School of Chemistry, Sun Yat-sen University, Guangzhou, China}

\author{\firstname{Duo} \surname{Zhang}\orcidlink{0000-0001-9591-2659}}
\affiliation{AI for Science Institute, Beijing, China}
\affiliation{Academy for Advanced Interdisciplinary Studies, Peking University, Beijing, China}

\author{\firstname{Chengqian} \surname{Zhang}}
\affiliation{AI for Science Institute, Beijing, China}
\affiliation{Academy for Advanced Interdisciplinary Studies, Peking University, Beijing, China}

\author{\firstname{Wanrun} \surname{Jiang}\orcidlink{0000-0001-7833-9862}}
\affiliation{AI for Science Institute, Beijing, China}

\author{\firstname{Yinan} \surname{Wang}}
\affiliation{AI for Science Institute, Beijing, China}

\author{\firstname{Antoine} \surname{Loew}\orcidlink{0009-0008-5018-4895}}
\affiliation{Ruhr University Bochum, Bochum, Germany}

\author{\firstname{Chengkun} \surname{Wu}\corres{}\orcidlink{0000-0002-9688-5311}}
\affiliation{College of Computer Science and Technology, National University of Defense Technology, Changsha, 410073, Hunan, China}

\author{\firstname{Weinan} \surname{E}\corres{}\orcidlink{0000-0003-0272-9500}}
\affiliation{AI for Science Institute, Beijing, China}
\affiliation{Center for Machine Learning Research, Peking University, Beijing 100871, P.R.China}
\affiliation{School of Mathematical Sciences, Peking University, Beijing, 100871, P.R.China}

\author{\firstname{Linfeng} \surname{Zhang}\corres{}\orcidlink{0000-0002-8470-5846}}\email{zhanglf@aisi.ac.cn}
\affiliation{AI for Science Institute, Beijing, China}
\affiliation{DP Technology, Beijing, China}

\author{\firstname{Han} \surname{Wang}\corres{}\orcidlink{0000-0001-5623-1148}}\email{wang\_han@iapcm.ac.cn}
\affiliation{National Key Laboratory of Computational Physics,
  Institute of Applied Physics and Computational Mathematics, Beijing, China}
\affiliation{HEDPS, CAPT, College of Engineering, Peking University, Beijing, China}



\begin{abstract}

Large Atomistic Models (LAMs) have undergone remarkable progress recently, emerging as universal or fundamental representations of the potential energy surface defined by the first-principles calculations of atomistic systems.
However, our understanding of the extent to which these models achieve true universality, as well as their comparative performance across different models, remains limited. 
This gap is largely due to the lack of comprehensive benchmarks capable of evaluating the effectiveness of LAMs as approximations to the universal potential energy surface.
In this study, we introduce LAMBench, a benchmarking system designed to evaluate LAMs in terms of their generalizability, adaptability, and applicability.
These attributes are crucial for deploying LAMs as ready-to-use tools across a diverse array of scientific discovery contexts.
We benchmark ten state-of-the-art LAMs released prior to August 1, 2025, using LAMBench. 
Our findings reveal a significant gap between the current LAMs and the ideal universal potential energy surface.
They also highlight the need for incorporating cross-domain training data, supporting multi-fidelity modeling, and ensuring the models' conservativeness and differentiability.
As a dynamic and extensible platform, LAMBench is intended to continuously evolve, thereby facilitating the development of robust and generalizable LAMs capable of significantly advancing scientific research.
The LAMBench code is open-sourced at \url{https://github.com/deepmodeling/lambench}, and an interactive leaderboard is available at
\url{https://www.aissquare.com/openlam?tab=Benchmark}.
\end{abstract}
\maketitle

\section{Introduction}\label{sec1}

The widespread adoption of large language models (LLMs) is largely driven by the development of general-purpose foundation models pretrained on vast and diverse corpora covering a wide range of disciplines and topics\cite{naveed2024comprehensiveoverviewlargelanguage}.
These foundation models are feasible because there exist common patterns to learn — namely, the shared logic of human language — despite its apparent diversity. 
In the field of molecular modeling, the fundamental physical principles of quantum mechanics, particularly the Schrödinger equation\cite{schrodinger1926quantisierung}, apply universally to all atomistic systems, assuming that relativistic effects are negligible.
Under the Born-Oppenheimer approximation\cite{born1985quantentheorie} a universal potential energy surface (PES) is defined as the ground state solution of the electronic Schrödinger equation, with the nuclear positions treated as input parameters.
Consequently, it is feasible to develop a foundational machine learning model to approximate this universal PES.
We refer to these molecular foundation models as Large Atomistic Models (LAMs) to emphasize their role in capturing fundamental atomic and molecular interactions across diverse chemical systems\cite{zhang2024dpa}.
LAMs are typically developed through a two-stage process: an initial pretraining phase on broad, diverse atomic datasets to learn a latent representation of the  universal PES, 
followed by fine-tuning on specific downstream datasets to specialize the model for particular target applications.

Despite the existence of a universal solution to the electronic Schrödinger equation, solving it remains computationally demanding even with modern quantum Monte Carlo methods\cite{austin2012quantum}.
In practice, Kohn-Sham density functional theory (DFT)\cite{hohenberg1964inhomogeneous}\cite{kohn1965self} is the most widely employed computational method for approximating the Born-Oppenheimer PES. 
The accuracy of DFT calculations is heavily contingent upon the modeling of the exchange-correlation functional, which varies across different research domains.
For instance, in materials science, the PBE/PBE+U\cite{perdew1996generalized} generalized gradient approximation (GGA) functionals are typically adequate, whereas in chemical science, GGA functionals often fall short, necessitating the use of hybrid functionals\cite{becke1993new} for improved accuracy\cite{mardirossian2017thirty}.
The disparity in exchange-correlation functionals, along with variations in the choice of basis sets and pseudopotentials, prevents the merging of DFT data across different research domains, thereby impeding the training of a universal potential model.

Nevertheless, domain-specific LAMs are advancing rapidly. For example, MACE-MP-0\cite{batatia2023foundation} and SevenNet-0\cite{park2024scalable} are both trained on the MPtrj dataset\cite{Deng2023} from the \textit{Inorganic Materials} domain at the 
PBE/PBE+U level of theory.
AIMNet\cite{doi:10.1126/sciadv.aav6490} and Nutmeg\cite{eastman2024nutmegspicemodelsdata} are trained in the domain of small molecules at the SMD(Water)-$\omega$B97X/def2-TZVPP {and the $\omega$B97M-
D3(BJ)/def2-TZVPPD level of theory, respectively}.
The rapid advancement of these domain-specific LAMs has transformed the field of atomistic modeling, offering powerful tools for understanding complex inorganic materials and bio-molecular systems. 
To fully harness the diverse training data from various research domains and maximize the potential of LAMs in learning universal PES, the multitask pretraining strategy presents a promising approach. 
This strategy encodes shared knowledge into a unified structure
with high representational capacity while integrating domain-specific components into multiple neural networks with relatively lower representational power\cite{shoghi2024moleculesmaterialspretraininglarge,zhang2024dpa,wood2025umafamilyuniversalmodels}.
However, determining the extent to which multitask-trained LAMs approach a truly universal PES remains a challenging question.

Comprehensive and robust benchmarking has proven to be a fundamental prerequisite for the rapid advancement of large-scale machine learning models.
For example, benchmarks such as MMLU-Pro\cite{wang2024mmluprorobustchallengingmultitask} and MATH500\cite{lightman2023lets} have driven the rapid progress of LLMs, while the ImageNet\cite{imagenet_cvpr09} benchmark has spurred the rapid iteration of modern computer vision models. Similarly, the CASP benchmark\cite{kryshtafovych2019critical} has played a crucial role in advancing protein structure prediction, ultimately leading to the development of AlphaFold2\cite{jumper2021highly}. 

In the field of molecular modeling, existing benchmarks exhibit two significant limitations. 
Firstly, they are intrinsically domain-specific, concentrating on isolated sub-fields rather than encompassing a variety of atomistic systems. 
For instance, datasets such as QM9\cite{ramakrishnan2014quantum} and MD17\cite{chmiela2017machine} are used to benchmark molecular property predictions and molecular dynamics (MD) trajectories of small molecules, respectively. 
These benchmarks are predominantly employed to assess machine learning models within chemical science.
The Matbench Discovery\cite{riebesell2024matbenchdiscoveryframework} evaluates models in the \domainim\ domain based on their ability to predict material stability. 
The Open Catalyst challenges\cite{ocp_dataset} assess models on predicting adsorption energies and relaxed structures for various adsorbate-catalyst combinations. 
While these benchmarks have played a crucial role in advancing domain-specific LAMs, their fragmented approach undermines the pursuit towards the universal PES model.
Secondly, existing assessment methods often fail to reflect real-world application scenarios, reducing their relevance to scientific discovery and technological innovation. 
For instance, conventional evaluation metrics based on static test sets may not adequately capture the true performance of a model in tasks requiring physically meaningful energy landscapes\cite{fu2025learningsmoothexpressiveinteratomic}.
Specifically, non-conservative models -- where atomic forces are directly inferred from neural networks rather than obtained from the gradient of the predicted energy\cite{neumann2024orb} -- can exhibit high apparent accuracy but struggle in applications demanding strict energy conservation, such as MD simulations\cite{fu2023forces}. 
The MLIP-Arena benchmark\cite{chiang2025mlip} is a step in the direction toward bridging this gap, emphasizing the practical usability of LAMs in tasks such as MD stability and physical property predictions.
However, it places less emphasis on {evaluating a model’s capacity to generalize across diverse atomistic systems or adapt to tasks beyond its training scope}, both of which are essential for assessing the performance of LAMs in real scientific discovery.

To address these limitations, we introduce LAMBench, a comprehensive benchmark system designed to rigorously evaluate LAMs across domains, simulation regimes, and application scenarios. 
Utilizing LAMBench, we assessed the performance of ten leading LAMs released before August 1, 2025, excluding the UMA models\cite{wood2025umafamilyuniversalmodels} due to licensing restrictions\cite{uma-license}. 
Our analysis revealed a substantial discrepancy between these models and the universal PES.
Our findings suggest that enhancing LAM performance requires simultaneous training with data from a diverse array of research domains. 
Additionally, supporting multi-fidelity at inference time is essential to satisfy the varying requirements of exchange-correlation functionals across different domains.
It is also critical to maintain the model's conservativeness and differentiability to optimize performance in property prediction tasks and ensure stability in molecular dynamics simulations.
We believe that the introduction of LAMBench will significantly expedite the development of LAMs, facilitating the creation of ready-to-use models that enhance the pace of real scientific discovery.

\section{Results}\label{sec2}

\subsection{The LAMBench system}
\label{sec:lambench-system}

\begin{figure}
    \centering
    \includegraphics[width=0.98\linewidth]{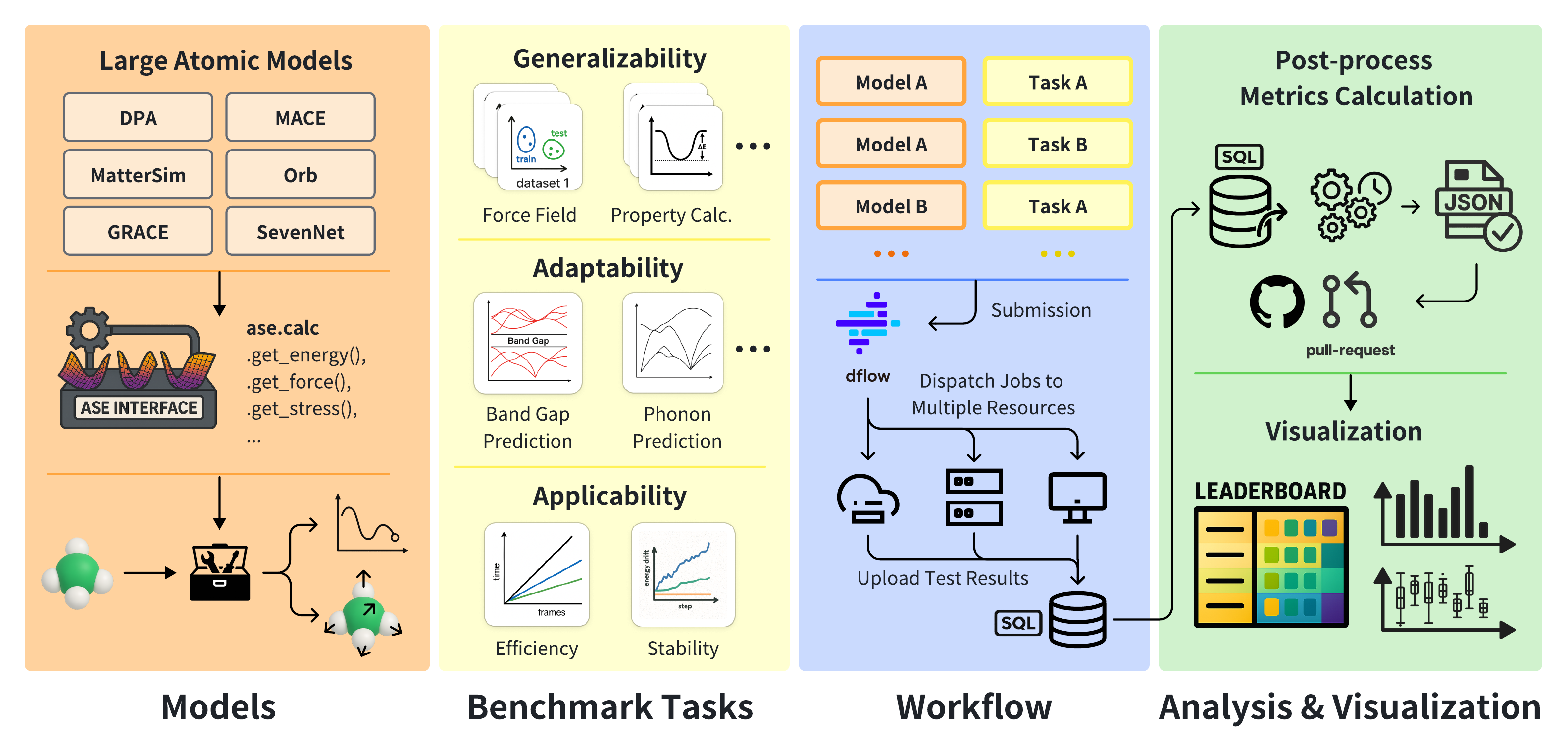}
    \caption{The schematic plot of the LAMBench benchmark.}
    \label{fig:toolkit}
\end{figure}

The LAMBench system is designed to benchmark diverse Large Atomistic Models (LAMs) across multiple tasks within a high-throughput workflow, with automation integral to task calculation, result aggregation, analysis, and visualization, as depicted in \cref{fig:toolkit}. 
The implementation details of LAMBench are elaborated in \cref{sec:implementation}.
Key to the system's effectiveness are the design of the benchmark tasks and the methodologies employed for result interpretation.
These benchmark tasks are developed to assess three fundamental capabilities of an LAM: generalizability, adaptability, and applicability.
Generalizability pertains to the accuracy of an LAM when utilized as a universal potential across a diverse range of atomistic systems. 
Adaptability denotes the LAM's capacity to be fine-tuned for tasks beyond potential energy prediction, with particular emphasis on structure-property relationship tasks. 
Applicability, on the other hand, concerns the stability and efficiency of deploying LAMs in real-world simulations.
In this study, all tasks are defined, and all results are obtained using LAMBench version v0.3.1.

Generalizability refers to the accuracy of an LAM on datasets that are not included in the training set.
In-distribution (ID) generalizability specifically pertains to the model's performance on test datasets generated through random splitting from the training datasets, thereby ensuring that these datasets maintain the same distribution as the training data.
Conversely, out-of-distribution (OOD) generalizability assesses the model's performance on test datasets independently constructed, resulting in a distribution distinct from the training data.
Within the LAMBench framework, references to generalizability always imply OOD generalizability.

It is important to note that there remains no consensus on the precise definition of OOD or the criteria by which two distributions are considered different. 
Some researchers define OOD data as those exploring different configurational spaces\cite{Morrow_2023,Deng2025}, while others highlight differences in chemical space as critical\cite{ocp_dataset,li2024probingoutofdistributiongeneralizationmachine}.
In this study, we adopt a practical approach by considering OOD test datasets as downstream datasets designed to address specific scientific challenges\cite{Mazitov_2024,PhysRevMaterials.7.045802,Ca_batteries_CM2021,doi:10.1021/acs.jpcc.2c08429,Batzner2022,Gao2025,10.1063/1.5023802,doi:10.1126/sciadv.adf0873,Wang2023,doi:10.1021/acs.jpclett.9b00085,Vandermause2022,doi:10.1021/jacs.3c10685}. 
These scenarios are most compatible with the downstream applications of LAMs.

In LAMBench, the generalizability is quantitatively assessed using 
two types of tasks, 
the force field task and the property calculation task. 
The force field task measures the accuracy of LAMs in predicting energy, force, and the virial tensor (under periodic boundary conditions) using twelve downstream datasets employed in the literature to train machine learning potentials for solving scientific problem, without any fine-tuning.
Detailed information on these datasets is provided in \cref{tab:datasets}. 
These test datasets encompass 
three distinct scientific domains: \domainim, \domaincat, and \domainsm.

It is important to note that the majority of LAMs are trained on datasets labeled with PBE/PBE+U exchange-correlation (XC) functionals, which may not align with the XC functional used in the test datasets.
To address this mismatch when evaluating the generalizability of LAMs, we relabeled five datasets at the PBE level, as detailed in the Methods section \cref{sec:methods}. 
Dispersion corrections were added through post-prediction process with the DFTD3-ASE interface\cite{10.1063/1.3382344, https://doi.org/10.1002/jcc.21759}, when necessary.
Minor mismatches in pseudopotentials, basis sets, and software implementations within DFT, which are less significant than the XC functional discrepancies, were not addressed in the current benchmark. 
Given that energy labels calculated via DFT can vary by an arbitrary constant,
LAMs are consistently used to predict the energy difference between the label and a dummy model that estimates potential energy solely based on the chemical formula. 
In contrast, force and virial predictions are directly obtained from the LAMs, as the dummy model invariably yields null predictions for force and virial.

Establishing a comprehensive performance indicator that synthesizes energy, force, and virial errors for each configuration across all datasets and domains presents a complex challenge.
An arithmetic mean of the error data is inappropriate due to the differing units of the quantities involved. 
To address this issue, we introduce a dimensionless error metric, $\bar{M}^m_{\mathrm{FF}}$, which essentially averages the relative errors in predictions across all datasets. 
Detailed information on this metric can be found in \cref{sec:generalizability-score}.
The  metric is structured so that a dummy model yields a value of $\bar{M}^m_{\mathrm{FF}}=1$, whereas an ideal model that perfectly aligns with DFT labels achieves a metric of $\bar{M}^m_{\mathrm{FF}}=0$.
A lower value of this error metric signifies enhanced generalizability.

In the error metric, the relative error is preferred over the absolute error to better reflect a model's accuracy in relation to the task's difficulty.
For example, the standard deviations in energy labels for the \textit{Lopanitsyna2023Modeling} and MD22 datasets are 511 meV/atom and 9 meV/atom, respectively. 
This considerable difference in scales indicates the greater diversity of configurations in the \textit{Lopanitsyna2023Modeling} dataset. 
If a model exhibits the same energy RMSE of 9 meV/atom for both datasets, it achieves a 2~\% error in the \textit{Lopanitsyna2023Modeling} dataset, while demonstrating no accuracy in the MD22 dataset.
Therefore, given the substantial differences between datasets, the relative error offers a more meaningful measure of model performance.

Achieving high accuracy in the force field task does not necessarily ensure enhanced performance in property calculations when utilizing the LAM as the force field\cite{fu2025learningsmoothexpressiveinteratomic}.
For instance, predicting the bulk and shear moduli necessitates finite difference approximations of the second-order derivatives of the potential energy. 
Consequently, models lacking smoothness up to the second-order derivatives may not consistently yield accurate predictions, even if they perform well in force field tasks.
Given the significant variation in investigated properties across different research domains, LAMBench offers a flexible and modular design (see \cref{sec:implementation}). 
This design facilitates the implementation of property calculation benchmark tasks, enabling them to be tailored to the specific requirements of diverse research fields.

In this study, we include property calculation tasks across the aforementioned research domains. 
For the \domainim\ domain, we adapt the MDR Phonon benchmark \cite{loew2024universal} to assess the performance of LAM models in computing essential phonon properties, including maximum phonon frequency ($\omega_{\max}$), entropy ($S$), free energy ($F$), and heat capacity at constant volume ($C_{V}$). 
Additionally, we employ the elasticity benchmark \cite{deJong2015, Liu_MatCalc_2024} to evaluate the accuracy of LAM models in calculating the shear modulus (GVHR) and bulk modulus (KVHR).
For the \domainsm\ domain, we use the TorsionNet500 benchmark\cite{Rai2020TorsionNet} to evaluate LAM performance in calculating torsion energy profiles of typical drug-like fragments, and the Wiggle150 benchmark\cite{Brew2025} to assess energy profiles of highly strained conformers.
For the \domaincat\ domain, we employ the OC20NEB-OOD dataset\cite{Wander2025} to benchmark the ability of LAMs to predict energy barriers for three reaction types—transfer, dissociation, and desorption—using DFT-optimized structures along the NEB trajectories.
For benchmarks in the \domainim\ domain, the PBE/PBE+U XC functional was adopted to compute the reference predictions. 
For the \domainsm\ and \domaincat\ domains, reference data are obtained at the CCSD(T) and RPBE levels, respectively, which are considered to be more accurate than the PBE XC functional within their respective domains.
The incorporation of these property calculation tasks serves as a prototype for assessing downstream generalizability in domain-specific property calculations. 
Additional tasks can be seamlessly integrated to align with the evolving capabilities of LAMs. 
In line with the approach used in the force field task, we introduce a dimensionless error metric, denoted as $\bar{M}^m_{\mathrm{PC}}$, to evaluate the generalizability of an LAM in property calculation tasks.
For comprehensive details, please refer to \cref{sec:generalizability-score}.

Adaptability assesses the ability with which a pre-trained LAM can be fine-tuned for tasks beyond its initial training scope, with a particular focus on establishing structure-property relationships.
This is distinct from the property calculation task. 
In the adaptability task, properties are directly predicted by fine-tuned LAMs, whereas, in the property calculation task, properties are computed using LAMs functioning as a force field.
The adaption provides a promising strategy for enhancing model accuracy in property prediction tasks, especially in situations where limited training data restricts the achievement of high accuracy through training from scratch\cite{shoghi2024moleculesmaterialspretraininglarge}. 
LAMs are typically pre-trained on force field prediction tasks\cite{zhang2024dpa,batatia2023foundation,park2024scalable} or on denoising tasks\cite{neumann2024orb}, which aim to recover stable configurations (local minima of the potential energy surface) from random perturbations in coordinates and atom type masking. 
Consequently, the ability of LAMs to adapt to property prediction tasks is not straightforward, highlighting the need for a benchmark to evaluate this aspect of LAMs.

Most LAMs comprise a feature extraction module, also referred to as a descriptor, and a fitting module. 
The feature extraction module encodes information about the universal PES into a latent space during pretraining, while the fitting module decodes this information to perform force field predictions. 
To predict a new downstream property, another fitting module is randomly initialized and fine-tuned jointly with the feature extraction module, while the original fitting module is discarded. 
Currently, adaptability tests for LAMs are exclusively supported for models implemented in the DeePMD-kit package\cite{zeng2025deepmdkitv3multiplebackendframework},  while other implementations are not supported. 

In this work, we use eight regression tasks from the MatBench benchmark\cite{dunn2020benchmarking} as representative examples. 
These tasks include the prediction of formation energies for crystal structures and perovskite cells, computed band gaps, exfoliation energies of two-dimensional materials, maximum phonon frequencies of bulk crystalline materials, dielectric constants, and shear moduli. 
Each task is evaluated using five-fold cross-validation. All properties are treated as intensive quantities, with mean pooling applied to atom-wise predictions.
Other tasks, including small molecule property predictions, spectroscopic property predictions, and classification tasks, can be readily integrated into the LAMBench framework in the future.

Finally, applicability assesses the readiness of LAMs for real-world deployment, focusing on computational efficiency and stability. 
Efficiency typically refers to the time required to compute energy, force, and virial (when applicable) using an LAM on a specific type of computational hardware. 
Stability examines whether the total energy of a system remains bounded, rather than diverging, during long-timescale MD simulations. 

Quantifying the inference efficiency of LAMs in a rigorous and meaningful manner requires careful design, as the observed efficiency can be highly structure-dependent, and different LAMs may exhibit varying performance on the same input. 
LAMs are designed to be applicable to tasks involving systems of varying sizes, from evaluating the conformational energy of molecules composed of dozens of atoms to large-scale simulations of viruses involving millions of atoms\cite{Lynch2023}. 
The efficiency of LAMs—measured as the computational time consumed per atom while determining the energy, forces, and virial of atomistic systems—is significantly influenced by the system size and the extent to which the many-core parallelism of modern GPUs is utilized. 
For relatively small systems, atom-wise efficiency is often reduced due to limited opportunities for parallelization, even though they require less overall computational time for evaluation.
As system size increases to approximately 1000 atoms, the average inference time per atom tends to stabilize, as illustrated by Supporting Information \cref{fig:si-eff_converge}.
For substantially larger systems, exceeding the memory capacity of a single GPU, multi-GPU parallelism becomes essential.
However, such systems are not ideal benchmarks for assessing the efficiency of LAMs, as they are generally computationally intensive, and multi-GPU parallelism is not supported by all the LAMs.

To address the significant size-dependency issues when measuring efficiency in small systems, we concentrate on atomistic systems that are large enough to achieve converged efficiency yet small enough to be computed without the need for multi-GPU parallelism. 
To this end, we randomly sample 1,000 structures from the \domainim\ and \domaincat\ domains in the aforementioned force field task test sets.
Each structure is subjected to periodic boundary conditions and is replicated by a multiplier, dynamically determined using a binary search algorithm. 
This approach, with an upper limit of 1,000 atoms for efficiency, ensures that measurements are conducted in or near the convergence regime. 
The \domainsm\ domain is excluded from this evaluation, as its systems contain only a few dozen atoms and therefore do not impose sufficient computational demands to meaningfully stress the inference capabilities of LAMs.

To facilitate the comparison of efficiency across models, we propose a dimensionless efficiency metric, denoted as $M_{\mathrm{E}}^m$. 
This metric is defined as the normalized inverse of the average inference time measured for 900 out of the 1,000 randomly sampled structures.
A detailed definition is provided in \cref{sec:applicability-score}. 
A value of $M_{\mathrm{E}}^m=1$ corresponds to an efficiency equivalent to a reference value of 0.01~$(\mathrm{\mu s/atom})^{-1}$, with higher value indicating greater efficiency. 
It is important to note that the calculator is fully warmed up by processing 100 structures, which are excluded from the 1,000 randomly sampled structures, before recording the inference time on the remaining 900 structures.

We assessed the stability of LAMs by examining energy conservation in NVE ensemble (microcanonical ensemble) MD simulations across nine atomistic systems. 
The extent of drift observed over time serves as a practical indicator of a model's stability, specifically in maintaining a finite system energy in extended MD trajectories.
These test systems were randomly selected from diverse domains, including four periodic structures from the Materials Project\cite{Jain2013}, three molecular systems from the SPICE2 dataset\cite{eastman_2024_10975225}, and two catalytic surface systems from the OC2M dataset\cite{zitnick2020introductionelectrocatalystdesignusing}, as detailed in Supporting Information \cref{fig:si-nve_struct}. 
For each system, a 10~ps NVE-MD simulation was conducted with a timestep of 1~fs. 
The initial atomic velocities were sampled from the Boltzmann ensemble at 300 K. 
The drift of total energy along MD trajectories was quantified using the slope derived from a linear regression applied to the total energy history, with the initial 2 ps of simulation excluded as a warm-up period.
Stability is assessed via the instability metric, $M^m_{\mathrm{IS}}$, which is calculated based on total energy drift, where smaller values denote improved stability.
A detailed definition is provided in \cref{sec:applicability-score}.

\setlength{\LTcapwidth}{\textwidth} 
\newcolumntype{L}[1]{>{\scriptsize\raggedright\arraybackslash}p{#1}}
\begin{longtable}{L{4.2cm} L{1.8cm} L{1.4cm} L{1.4cm} L{5cm}}
    \caption{Summary of datasets utilized in force field generalizability assessments.
    The table details the domain of each test dataset, available labels (E for energy, F for force, and V for virial), number of configurations (frames), exchange-correlation (XC) functional applied for labeling, and additional descriptions.
    } \label{tab:datasets} \\

    \toprule
    \textbf{Name} & \textbf{Domain} & \textbf{Labels} & \textbf{Frames} & \textbf{Description} \\
    \midrule
    \endfirsthead
    
    \multicolumn{5}{r}{\textit{Continued from the previous page}} \\
    \toprule
    \textbf{Name} & \textbf{Domain} & \textbf{Labels} & \textbf{Frames} & \textbf{Description} \\
    \midrule
    \endhead
    
    \midrule
    \multicolumn{5}{r}{\textit{Continued on the next page}} \\
    \endfoot
    \bottomrule
    \multicolumn{5}{l}{} \\
    \endlastfoot
        \textit{Mazitov2024Surface}\cite{Mazitov_2024} & Inorganic Materials & EFV &422 & A down-sampled dataset of \textit{HEA25S}, for high-entropy alloy surfaces, focusing on d-block transition metals covering 25 elements. \\
        \addlinespace
        \textit{Lopanitsyna2023Modeling}\cite{PhysRevMaterials.7.045802} & Inorganic Materials & EFV & 900 & A down-sampled dataset of \textit{HEA25}, for high-entropy alloy bulk structures, focusing on d-block transition metals covering 25 elements. \\
        \addlinespace
        \textit{Torres2019Analysis}\cite{Ca_batteries_CM2021} & Inorganic Materials & EF & 1000 & A down-sampled dataset of Ca-bearing minerals, focusing on silicates and carbonates. \\
        \addlinespace
        \textit{Sours2023Predicting}\cite{doi:10.1021/acs.jpcc.2c08429} & Inorganic Materials & EF & 1000 & A down-sampled Dataset consisting of 219 different pure silica zeolite topologies. \\
        \addlinespace
        \textit{Batzner2022equivariant}\cite{Batzner2022} & Inorganic Materials & EF & 1000 & A down-sampled dataset containing lithium phosphate amorphous glass used by NequIP graph neural network models as evaluation configurations. \\
        \addlinespace
        
        \textit{Gao2025Spontaneous}\cite{Gao2025} & Inorganic Materials & EFV & 1000 & A down-sampled Bilayer 2D MoS$_2$ structures. \\
        \addlinespace
       
        \textit{ANI-1x}\cite{10.1063/1.5023802} & Molecules & EF & 997 & A down-sampled dataset from the training data of the ANI-1x potential, containing organic molecule structures. \\
        \addlinespace
        
        \textit{MD22}\cite{doi:10.1126/sciadv.adf0873} & Molecules & EF & 1000 & A down-sampled dataset from MD22, where molecular dynamics (MD) trajectories of 42-atom tetrapeptide Ac-Ala3-NHMe, docosahexaenoic acid, Stachyose, DNA base pair (AT-AT), DNA base pair (AT-AT-CG-CG). \\
        \addlinespace
        \textit{AIMD-Chig}\cite{Wang2023} & Molecules & EF & 942 & A down-sampled MD dataset containing conformations of chignolin protein.\\
        \addlinespace

        \textit{Zhang2019Bridging}\cite{doi:10.1021/acs.jpclett.9b00085} & Catalysis & EF & 1000 & A down-sampled dataset for interaction of carbon dioxide with a movable Ni(100) surface. \\
        \addlinespace
        \textit{Vandermause2022Active}\cite{Vandermause2022} & Catalysis & EFV & 250 & A down-sampled dataset for direct simulation of hydrogen turnover on Pt(111) catalyst surfaces. \\
        \addlinespace
        \textit{Villanueva2024Water}\cite{doi:10.1021/jacs.3c10685} & Catalysis & EF & 1000 & A down-sampled dataset for selective CO$_2$ hydrogenation to methanol over oxide catalysts. \\
 \\
  
\end{longtable}

\subsection{Benchmark results}

\Cref{tab:lambench} presents the LAMBench Leaderboard, showcasing the performance of LAMs in terms of their generalizability and applicability.
Currently, adaptability is excluded from the comparison due to  the absence of this test for most of the LAMs under evaluation. 
This leaderboard aims to strike a balanced comparison between key performance aspects and is intended to offer insights into model suitability for applications relevant to real-world scientific discovery.
To reduce discrepancies in the comparisons between multi-task and single-task trained models, we employed the MPtrj task head for the DPA-3.1-3M and DPA-2.4-7M models, and the MPA modality for the SevenNet-MF-ompa model, unless stated otherwise.

\begin{table}
    \caption{The LAMBench Leaderboard.
    Details regarding the calculation of leaderboard metrics are provided in \cref{sec:methods}. $\bar M^m_{\mathrm{FF}}$ refers to the generalizability error on force field prediction tasks, while $\bar M^m_{\mathrm{PC}}$ denotes the generalizability error on domain-specific tasks. 
    $M_{\mathrm{E}}^m$ stands for the efficiency metric, and $M^m_{\mathrm{IS}}$ refers to the instability metric. Arrows alongside the metrics denote whether a higher or lower value corresponds to better performance.}
    \centering
    \begin{tabular}{@{}lw{c}{1.8cm}w{c}{1.8cm}w{c}{1.8cm}w{c}{1.8cm}@{}}
    \toprule
        \multirow{2}{*}{\textbf{Model}} & \multicolumn{2}{c}{\textbf{Generalizability}} & \multicolumn{2}{c}{\textbf{Applicability}} \\
        \cmidrule(lr){2-3} \cmidrule(lr){4-5}
         & $\bar M^m_{\mathrm{FF}} \downarrow $& $\bar M^m_{\mathrm{PC}} \downarrow $& $M_{\mathrm{E}}^m  \uparrow$ & $M^m_{\mathrm{IS}} \downarrow $\\
        \midrule
        DPA-3.1-3M         & \normcolor{0.175}{0.175}{0.351} & \normcolor{0.322}{0.322}{0.601}  & \normcolor{0.261}{1.341}{0.084} & \normcolor{0.572}{0.0}{2.649}\\
        Orb-v3        & \normcolor{0.215}{0.175}{0.351} & \normcolor{0.414}{0.322}{0.601}  & \normcolor{0.396}{1.341}{0.084} & \normcolor{0.000}{0.0}{2.649}\\
        DPA-2.4-7M         & \normcolor{0.241}{0.175}{0.351} & \normcolor{0.342}{0.322}{0.601}  & \normcolor{0.617}{1.341}{0.084} & \normcolor{0.039}{0.0}{2.649}\\
        GRACE-2L-OAM       & \normcolor{0.251}{0.175}{0.351} & \normcolor{0.404}{0.322}{0.601}  & \normcolor{0.639}{1.341}{0.084} & \normcolor{0.309}{0.0}{2.649}\\
        Orb-v2             & \normcolor{0.253}{0.175}{0.351} & \normcolor{0.601}{0.322}{0.601}  & \normcolor{1.341}{1.341}{0.084} & \normcolor{2.649}{0.0}{2.649}\\
        SevenNet-MF-ompa   & \normcolor{0.255}{0.175}{0.351} & \normcolor{0.455}{0.322}{0.601}  & \normcolor{0.084}{1.341}{0.084} & \normcolor{0.000}{0.0}{2.649}\\
        MatterSim-v1-5M    & \normcolor{0.283}{0.175}{0.351} & \normcolor{0.467}{0.322}{0.601}  & \normcolor{0.393}{1.341}{0.084} & \normcolor{0.000}{0.0}{2.649}\\
        MACE-MPA-0         & \normcolor{0.308}{0.175}{0.351} & \normcolor{0.425}{0.322}{0.601}  & \normcolor{0.293}{1.341}{0.084} & \normcolor{0.000}{0.0}{2.649}\\
        SevenNet-l3i5      & \normcolor{0.326}{0.175}{0.351} & \normcolor{0.397}{0.322}{0.601}  & \normcolor{0.272}{1.341}{0.084} & \normcolor{0.036}{0.0}{2.649}\\
        MACE-MP-0          & \normcolor{0.351}{0.175}{0.351} & \normcolor{0.472}{0.322}{0.601}  & \normcolor{0.296}{1.341}{0.084} & \normcolor{0.089}{0.0}{2.649}\\
        \bottomrule
    \end{tabular}
    \label{tab:lambench}
\end{table}

\begin{figure}
\centering
\includegraphics[width=1\linewidth]{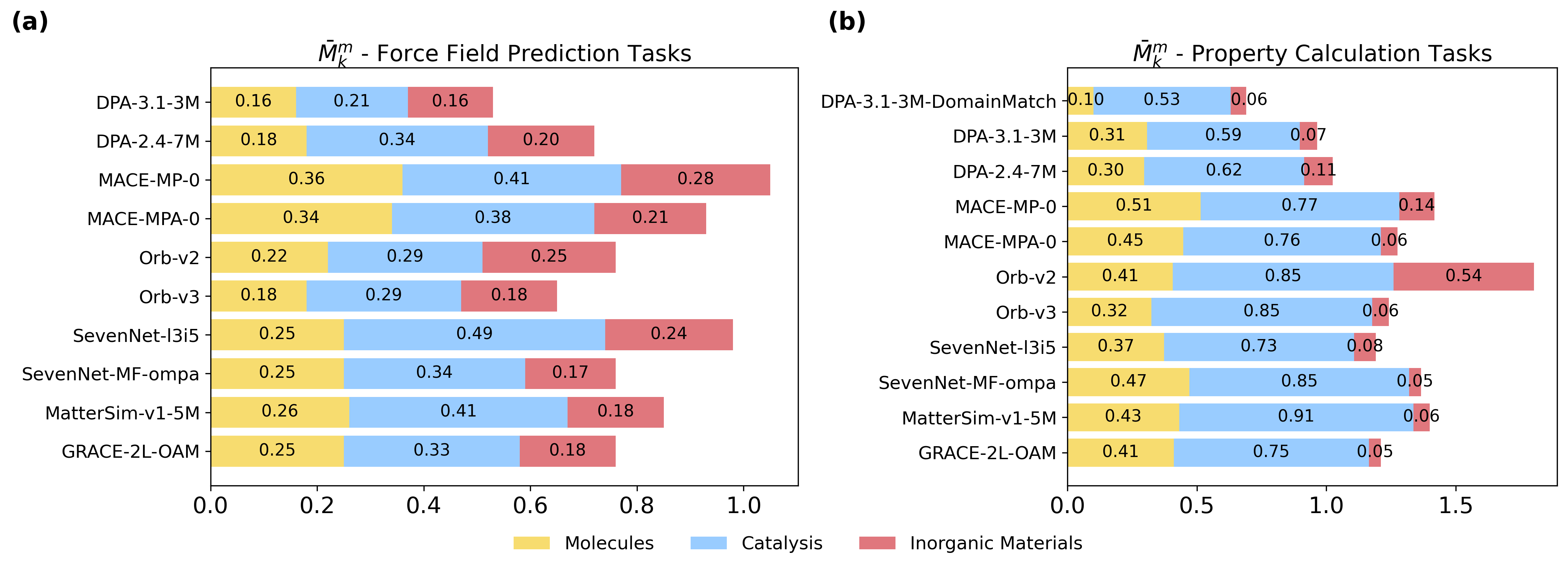}
\caption{Dimensionless error metrics for generalizability tasks across different domains. 
(a) Dimensionless error metrics for force field tasks across three distinct domains.
(b) Dimensionless error metrics for property-calculation tasks.
To reassess DPA-3.1-3M with enhanced domain alignment, referred to as DPA-3.1-3M-DomainMatch,  we employed task heads trained on \textit{SPICE2} for the \domainsm\ domain, \textit{OMAT24} for the \domainim\ domain, and \textit{OC20M} for the \domaincat\ domain.
}
    \label{fig:twobars}
\end{figure}

In the force field generalizability task, as illustrated in \cref{tab:lambench}, performance is evaluated using the dimensionless error $\bar{M}_{\mathrm{FF}}^m$. Notably, DPA-3.1-3M  demonstrates substantially greater generalizability compared to other LAMs. 



To further investigate the generalizability of these LAMs across specific domains, we present the domain-wise dimensionless error metric in \cref{fig:twobars}(a).
In general, most LAMs exhibit relatively low inference errors in the domains of \domainsm\ and \domainim, while showing inferior performance in the domain of \domaincat.
For \domainsm, DPA-3.1-3M ranks first, closely followed by DPA-2.4-7M and Orb-v3, and outperforms all other models in this domain.
In the \domainim\ domain, the comparative performance of the models shows a similar pattern to their rankings in the Matbench Discovery benchmark, accessed on August 1st, 2025. These models, excluding DPA2-2.4-7M, are ranked by their lowest MAE in the following order: SevenNet-MF-ompa $<$ GRACE-2L-OAM $<$ DPA-3.1-3M $<$ Orb-v3 $<$ MatterSim-v1-5M $<$ MACE-MPA-0 $<$ Orb-v2 $<$ SevenNet-l3i5 $<$ MACE-MP-0.
This coherence further substantiates the reliability and robustness of both the LAMBench and Matbench Discovery benchmark frameworks\cite{riebesell2024matbenchdiscoveryframework} in evaluating LAMs for \domainim\ force field tasks.
Within the \domaincat\ domain, DPA-3.1-3M again achieves the lowest error, significantly surpassing the other models.

\Cref{tab:lambench} indicates that MACE-MPA-0, which features an expanded set of parameters and is trained on a more comprehensive dataset, achieves a dimensionless error of 0.308. 
It is 0.043 lower than the error of MACE-MP-0, which stands at 0.351, suggesting improved generalizability. 
This finding is further supported by the domain-specific error metrics presented in \cref{fig:twobars}(a), where MACE-MPA-0 demonstrates superior performance across all three domains.
Notably, in the \domainim\ domain, there is an approximate 25\% reduction in error.
Similarly, the SevenNet-MF-ompa model shows a comparable improvement over its lighter version, SevenNet-l3i5.

The DPA-3.1-3M model\cite{zhang2025graphneuralnetworkera} demonstrates the best overall performance across all investigated domains. 
This achievement is likely due to its multi-task training strategy\cite{shoghi2024moleculesmaterialspretraininglarge, zhang2024dpa, wood2025umafamilyuniversalmodels} that train the model on 31 datasets spanning multiple domains, detailed in Supporting Information \cref{tab:si-datasets}. 
The model's outstanding performance in the \domainsm\ and \domaincat\ domains can be attributed to the descriptor, which was trained using diverse datasets from domains such as \domainsm\ (including the SPICE2 and Yang2023ab datasets) and \domaincat\ (including the OC20 and OC22 datasets).
The DPA-2.4-7M model, trained using the same multitask strategy, ranks third, showing a noticeable gap from DPA-3.1-3M and Orb-v3, largely due to its limited capacity.



The generalizability of LAMs in property calculation tasks is assessed and summarized in \cref{tab:lambench}.
Compared to force field tasks, property calculation tasks offer greater discriminative power among models. 
The DPA-3.1-3M model again achieves the best performance, with the lowest error of 0.322—representing nearly a fifty percent reduction in error relative to the worst-performing model.
The domain-wise error metrics are presented in \cref{fig:twobars}(b), with the detailed raw measurements available in Supporting Information \cref{tab:si-phonon,tab:si-elasticity,tab:si-torsion500,tab:si-wiggle,tab:si-neb}.

In the \domainim\ domain, where the MDR phonon and elasticity benchmarks are conducted, the conservative models exhibit significantly lower errors compared to the non-conservative Orb-v2 model, which directly predicts force.
Accurate phonon and elasticity predictions necessitate the calculation of the force constant matrix, representing the second-order derivative of a force field. 
Therefore, conservativeness and smoothness are crucial for achieving precise predictions, consistent with previous findings in Ref.\cite{fu2025learningsmoothexpressiveinteratomic}.
Additionally, the performance difference between MACE-MP-0 (0.14) and its more expressive counterpart, MACE-MPA-0 (0.06), suggests that enhanced generalizability in force field prediction tasks can lead to improved performance in a property prediction task. 
In both benchmarks, the Omat24 task head of DPA-3.1-3M, denoted as DPA-3.1-3M-DomainMatch, demonstrates improved performance over the MPtrj task head, underscoring the benefits of more extensive training data.

In the \domainsm\ domain, evaluations are conducted using the TorsionNet500 and Wiggle150 benchmarks.
The multitask-trained LAMs, specifically DPA-2.4-7M and DPA-3.1-3M, achieved the lowest errors of 0.30 and 0.31, respectively. 
These models outperformed others primarily trained on \domainim\ domain datasets, with the exception of Orb-v3, which unexpectedly achieved an error of 0.32. 
This performance, despite Orb-v3 being trained solely on \domainim\ domain datasets, may be attributed to its denoising pretraining strategy and conservative design. 
It is notable that the dimensionless errors observed in this domain were significantly higher compared to those seen in the \domainim\ domain.
This increase in error can be primarily attributed to two factors. 
First, there is a mismatch in XC functionals: predictions made by LAMs at the PBE level are less accurate compared to the gold-standard CCSD(T) references used in the benchmark.
When the domain-matched task head, specifically the SPICE2 head with the $\omega$B97MD3(BJ) XC functional, is used for property prediction, the error decreases from 0.31 (DPA-3.1-3M) to 0.10 (DPA-3.1-3M-DomainMatch), as illustrated in \cref{fig:twobars}.
Second, most LAMs, with the exception of the DPA models, have not been exposed to molecular systems during training, making cross-domain generalization inherently more challenging than in-domain generalization.
With the recent release of the OMol25 dataset \cite{levine2025openmolecules2025omol25}, we anticipate that LAMs, such as the UMA models, will more readily achieve excellent performance in the \domainsm\ domain.
Despite these achievements, LAMs still lag behind domain-specific approaches in property prediction.
For example, MACE-OFF23\cite{kovacs2023maceoff23} achieves a dimensionless error of 0.04, representing a 60\% reduction compared to the best-performing LAM, DPA-3.1-3M-DomainMatch. 

In the \domaincat\ domain, where the OC20NEB-OOD benchmark is conducted, all LAMs exhibit limited generalizability.
The top-performing model, DPA-3.1-3M-DomainMatch, utilizing the OC20M task head, still presents a relatively high dimensionless error of 0.53. In contrast, domain-specific models such as EquiformerV2-31M-S2EF-OC20-All+MD\cite{liao2024equiformerv2improvedequivarianttransformer,chanussot2021open} and eSCN-L6-M3-Lay20-S2EF-OC20-All+MD\cite{passaro2023reducingso3convolutionsso2,chanussot2021open} achieve substantially lower dimensionless errors of 0.31 and 0.33, respectively.
As illustrated in \cref{tab:si-neb}, LAMs achieve notably lower errors when predicting the reaction energy (\(dE\)) as opposed to the reaction barrier (\(E_a\)).
This discrepancy indicates that improving the performance of LAMs necessitates increased focus on the transition states.

The adaptation of the pretrained DPA-3.1-3M and DPA-2.4-7M models across eight Matbench property regression tasks is detailed in \cref{tab:property}. 
The fine-tuned model consistently surpasses the model with the same architecture trained from scratch in terms of accuracy.
Notably, a property predictor adapted from the pretrained MatterSim\cite{yang2024mattersimdeeplearningatomistic} achieves performance comparable to leading task-specific models on the Matbench leaderboard.
Furthermore, JMP, pretrained on the OC20, OC22, ANI-1x, and Transition-1x datasets\cite{shoghi2024moleculesmaterialspretraininglarge}, significantly surpasses existing models, achieving state-of-the-art accuracy.
Although a performance gap remains between DPA-3.1-3M and the state-of-the-art (SOTA) results, its superior performance compared to DPA-2.4-7M demonstrates that better generalizability in force field prediction tasks also translates to improved adaptability in property fine-tuning tasks.
This underscores the critical importance of pretraining for achieving superior model accuracy in downstream property prediction tasks and reinforces the broader vision of LAMs as versatile, high-performing surrogates capable of addressing a wide range of scientific challenges.

\begin{table}
    \caption{Adaptability Test: The accuracy of property fine-tuning of DPA models across eight Matbench regression tasks. \protect\footnotemark[1]
    }
    \vspace{8pt}
    \resizebox{\textwidth}{!}{ 
    {\fontsize{8}{9.5}\selectfont
    \begin{tabular}{lccccccc}
        
        \toprule
        \textbf{Task (Unit)} & \textbf{\makecell{DPA-2.4-7M \\From scratch}} &  \textbf{\makecell{DPA-2.4-7M \\Finetune}} & 
        \textbf{\makecell{DPA-3.1-3M \\From scratch}} &
        \textbf{\makecell{DPA-3.1-3M \\Finetune}} & 
        \textbf{\makecell{Matbench\cite{dunn2020benchmarking} \\leaderboard}} &
        \textbf{MatterSim\cite{yang2024mattersimdeeplearningatomistic}} &\textbf{JMP\cite{shoghi2024moleculesmaterialspretraininglarge}} \\
        \midrule
        MP $E_{\text{form}}$ (meV/atom) & 31.1 & 24.7 & 24.2 & 13.9 & 17.0 &-& 10.1\\
        MP Gap (eV) & 0.285  & 0.261 & 0.255 & 0.147 & 0.156 & 0.129& 0.091 \\
        JDFT2D (meV/atom) & 46.93  & 32.23 & 41.80 & 31.29 & 33.19 & 32.76& 29.94 \\
        Phonons (cm$^{-1}$) & 41.10  & 29.05 & 41.41 & 25.13 & 28.76 & 26.02& 20.57 \\
        Dielectric (unitless) & 0.431 &  0.292 & 0.438 & 0.297 & 0.271 & 0.252& 0.249 \\
        Log KVRH ($\log_{10}\text{GPA}$) & 0.062 & 0.052 & 0.060 & 0.051 & 0.049 & 0.049 & 0.045 \\
        Log GVRH ($\log_{10}\text{GPA}$) & 0.079  &  0.066 & 0.079 & 0.061 & 0.067 & 0.061 & 0.059 \\
        Perovskites (eV/unitcell) & 0.061  &  0.047 & 0.043 & 0.035 & 0.027 & - & 0.026 \\
        \bottomrule
    \end{tabular}
    }
    }
    \label{tab:property}

\footnotemark[1]{\footnotesize Results were obtained using 4 * NVIDIA L4-24GB GPUs with the maximum permissible batch size. For the \textit{MP $E_{\text{form}}$} and \textit{MP Gap} tasks, the reported accuracy may not reflect the fully converged results due to insufficient training epochs.
}
\end{table}

High efficiency and robust stability are equally critical as model accuracy, especially in the context of high-throughput simulations.
The efficiency metric $M^m_{\mathrm{E}}$ is summarized in \cref{tab:lambench}.
Despite having the most parameters, Orb-v2 demonstrates the highest inference efficiency. 
This superior performance is likely attributed to its non-conservative design, where force predictions are generated from a separate prediction head rather than being derived from energy gradients.

Additionally, we observe that the efficiency is highly sensitive to the test structure for certain LAMs, as exemplified by the broad distribution of the SevenNet-l3i5 model. 
In contrast, models such as Orb-v2 and DPA-2.4-7M demonstrate relative insensitivity to structural variations.
To further investigate the origin of this behavior, a bilayer sodium 2D structure was selected as a representative cases. 
We systematically reduced the vacuum spacing by shortening the $c$-axis, and the results are summarized in Supporting Information \cref{fig:na-nnei}~(a).
For Orb-v2 and DPA-2.4-7M, the converged efficiency remained largely unaffected by changes in vacuum spacing. 
In contrast, other LAMs experienced a significant drop in efficiency as the vacuum was reduced. 
Once the vacuum reached a certain threshold, further increases had no impact on efficiency. 
Reducing the vacuum effectively increases the number of neighbors, suggesting that the average number of neighboring atoms within the cutoff radius is the key factor influencing inference efficiency in these models, as demonstrated in Supporting Information \cref{fig:na-nnei}~(b). 
In contrast, Orb-v2 and DPA-2.4-7M utilize a fixed maximum number of neighbors through padding, rendering them insensitive to such variations.

\begin{figure}
        \centering
        \includegraphics[width=.7\textwidth]{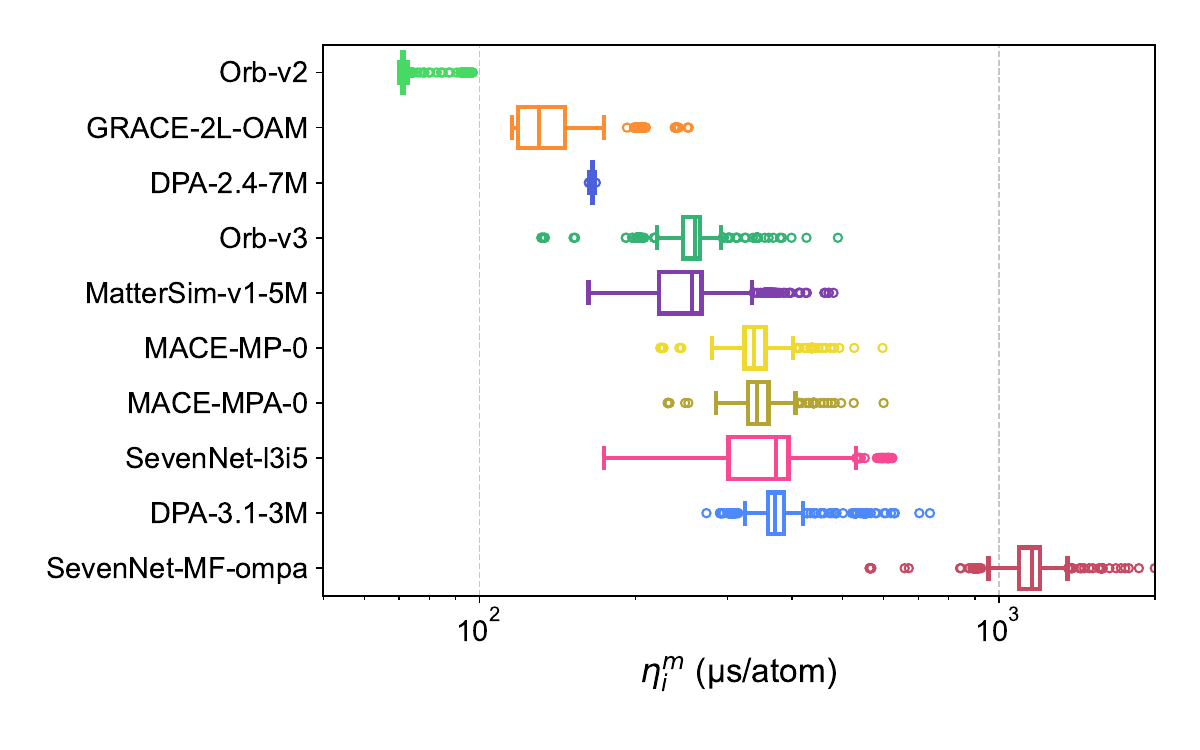}
        \caption{
       Distribution of inference time, normalized by the number of atoms, measured across 900 randomly selected configurations. Lower values indicate higher efficiency.
        }
        \label{fig:eff_vs_models}
\end{figure}

The stability of the LAMs is evaluated using the instability metric \( M^m_{\mathrm{IS}} \), as presented in \cref{tab:lambench}.
The conservative models, such as Orb-v3
, exhibit minimal energy drift throughout the evaluation, indicating stable simulations.
In contrast, the non-conservative model, Orb-v2, shows instability metric several orders of magnitude larger, reflecting instability in the simulation. 
The relatively high instability metric observed in DPA-3.1-3M is due to a failed simulation, potentially caused by the relatively low training weight assigned to the MPtrj task head, as detailed in \cref{tab:si-datasets}.
Notably, utilizing the Omat24 task head reduces the instability metric to zero. 
In general, models with a greater number of trainable parameters and trained on larger datasets exhibit higher stability, as reflected in reduced energy drift. 
This trend is evident in the MACE, and SevenNet model families as shown in Supporting Information \cref{tab:si-nvemd}.

\section{Discussion and Conclusion}

This study introduces a comprehensive benchmarking framework, designated as LAMBench, for evaluating large atomistic models (LAMs).
It is designed to assess the extent to which LAMs can be used as versatile, out-of-the-box tools capable of advancing scientific discovery across a broad range of contexts.
We emphasize three essential requirements for realizing this vision: generalizability to atomistic systems across diverse research domains, adaptability to novel tasks, and applicability in real-world simulations with regard to stability and efficiency.

We propose two test tasks to evaluate generalizability: the force field tasks, which assess LAMs' accuracy in predicting energy, force, and virial (when applicable), and the property calculation tasks, which measure accuracy in computing properties of interest for specific applications using the LAM as a force field. 
The errors of LAMs in these tasks are interpreted using a dimensionless error metric.
This metric facilitates the comparison of generalizability across different test tasks, despite significant variations in label magnitudes, and offers a clear indication of the model's proximity to an ideal universal model.


The LAMBench results highlight a substantial gap between current LAMs and the ideal universal model, envisioned as an out-of-the-box simulation tool for scientific discovery. 
This gap is primarily due to limited generalizability beyond the \domainim\ domain, which is largely attributed to training datasets predominantly sourced from this domain. 
More specialized datasets are needed to improve cross-domain generalization, such as those containing transition states in the \domaincat\ domain.
Although DPA-3.1-3M employs a multi-task training scheme, utilizing data from multiple domains to enhance its generalization capabilities and outperforming other single-task trained models in most situations, its accuracy remains inferior to models specifically tailored for individual domains.
To bridge this gap, future research might focus on developing advanced model architectures for improved generalizability, more efficient training methods to learn shared knowledge across diverse domains, and a balanced distribution of training data across various research fields. 
These directions hold promise for advancing the capabilities of LAMs in scientific discovery.

The capability to adjust model fidelity is proposed as an essential feature for LAMs, given that different research domains necessitate varying levels of accuracy in DFT calculations.
For example, the \domainim\ domain generally necessitates GGA functionals, whereas the \domainsm\ domain often requires hybrid or higher-level functionals. 
Training models across various XC functional fidelities offers considerable flexibility for downstream applications and should thus be actively promoted within the community \cite{kim_sevennet_mf_2024}.

Regarding the efficiency of using LAMs in simulations, the non-conservative model Orb-v2 demonstrates significantly higher performance compared to the conservative models. 
Among the conservative models, the fastest model is 7.6 times quicker than the slowest, a disparity that is notably larger than the variations observed in generalizability.
This observation underscores the importance of considering the efficiency in the development of LAMs, particularly for speed-critical applications such as long-time molecular dynamics simulations.

In terms of stability, conservative models exhibit significantly superior performance compared to non-conservative models, a finding that is corroborated by existing literature\cite{fu2025learningsmoothexpressiveinteratomic}.
This observation, along with the necessity for conservativeness in phonon calculation tasks, emphasizes the importance of integrating conservativeness into the design of LAMs.

In this study, adaptability is exclusively benchmarked for the DPA models, as most implementations of LAMs are primarily confined to force field prediction and are not readily adaptable to property prediction tasks.
The adaptability test for DPA-3.1-3M and DPA-2.4-7M demonstrates that a pretrained model can offer substantial advantages in downstream property prediction tasks compared to models trained from scratch, particularly in data-scarce scenarios. 
The comparison between these two models further suggests that greater generalizability in force field prediction tasks translates to improved adaptability in property fine-tuning tasks. 
Additional research is needed to validate this correlation across other LAMs. 
Furthermore, it is advisable for LAM developers to enhance their models with the capability to adapt for property prediction. 
This advancement could open up new avenues for their application in scientific discovery.

Currently, LAMBench only includes twelve OOD datasets for force field prediction and five for property calculation across three research domains, supporting eight adaptability and two applicability tests.
In the future, LAMBench should incorporate more comprehensive downstream test cases, encompassing both force field and property calculation scenarios, to more accurately reflect the performance of LAMs in real-world applications as an out-of-the-box simulation tool.
This enhancement necessitates dynamically adjusting the force field generalizability test datasets to accommodate emerging trends in training dataset development, which may render some OOD generalizability test cases in-distribution. 
Consequently, it will be essential to replace these test cases with more challenging alternatives.
It is important to note that the design of property calculation tasks requires careful consideration. 
Directly borrowing all property calculations from applications may not be optimal, as these calculations often necessitate long time-scale and large spatial scale MD simulations, such as for ionic diffusion constants, which demand substantial computational resources.
Ideally, property benchmarks should be designed as representative examples that reflect performance in the calculation of domain-specific properties while remaining computationally feasible.

Evaluating adaptability presents challenges, even when LAM implementations support fine-tuning for property prediction tasks. 
The adaptability benchmark necessitates a fine-tuning procedure, meaning that comparisons of final performance are influenced not only by the model's adaptability but also by variations in fine-tuning code. 
Establishing a unified property fine-tuning framework for all LAMs could be the most effective solution; however, this would require extensive development and is beyond the current scope of LAMBench.

Given the diverse application scenarios of LAMs, LAMBench is designed as a dynamic system, continuously incorporating additional test tasks and datasets over time. 
Establishing such a framework requires sustained, community-driven efforts and consensus; therefore, we strongly encourage ongoing contributions from the community.

\section{Method}

\label{sec:methods}

\subsection{computational details}
Energy and force labels of \textit{ANI-1x}, \textit{MD22} and \textit{AIMD-Chig} sets are calculated using density functional theory (DFT) at the level of generalized gradient approximation (GGA)  in the form of Perdew, Burke and Ernzerhof (PBE) functional\cite{PhysRevLett.77.3865}\cite{PhysRevLett.78.1396} with 6-31G(d) basis sets in Gaussian 16 Rev. C.01\cite{g16}.
self-consistent field (SCF) convergence requires the root mean squared (RMS) change in the density matrix smaller than $1 \times 10^{-8}$, meanwhile the maximum change in the density matrix smaller than $1 \times 10^{-6}$.
Integration grids are set to 99 radial shells per atom and 590 angular points per shell.
Calculations for \textit{ANI-1x} and \textit{MD22} systems adopt the neutral and closed-shell singlet states, which is consistent with the original datasets.
Calculations for \textit{AIMD-Chig} systems adopt the -2 charged and closed-shell singlet states, given all molecules in this set are chignolin dianions.


The energies, interatomic forces and virials labels of the \textit{Lopanitsyna2023Modeling} and \textit{Mazitov2024Surface} data on  were obtained from density functional theory (DFT) calculations performed using the Vienna Ab initio Simulation Package (VASP). 
The electronic structure calculations employed the generalized gradient approximation with the PBE functional for exchange-correlation effects, adopting parameter settings consistent with the Materials Project database, including a plane-wave cutoff energy of 520 eV. 
The electronic self-consistency cycle was converged to $5 \times 10^{-5}$ eV. 
For Brillouin zone integration, we used the Methfessel-Paxton smearing method (\texttt{ISMEAR} = 1) with a smearing width of 0.05 eV, which is particularly suitable for metallic systems like high-entropy alloys. 
The k-point mesh was generated automatically based on a reciprocal density of 64, corresponding to a k-spacing of approximately 0.5 $Å^{-1}$, ensuring sufficient sampling accuracy while maintaining computational efficiency. 
Core-electron interactions were treated using projector-augmented wave pseudopotentials, and all calculations were performed without spin polarization to maintain consistency across the diverse compositional space of our high-entropy alloy systems.

\subsection{Models}

To evaluate the thoroughness and discriminative power of the LAMBench benchmark, we test a series of LAMs, as listed in \cref{tab:models}. 
Most models follow a single-task, single-fidelity training strategy, using datasets curated under consistent DFT settings—primarily from the domain of \textit{Inorganic Materials}. Exceptions include SevenNet-MF-OMPA and DPA models.
SevenNet-MF-OMPA employs multi-fidelity training on the OMat24\cite{barrosoluque2024openmaterials2024omat24}, MPtrj, and sAlex\cite{barrosoluque2024openmaterials2024omat24} datasets, achieving high accuracy despite heterogeneity in DFT settings, though still focused on inorganic systems.
Conversely, both the DPA-3.1-3M and DPA-2.4-7M models implement a multitask training strategy utilizing the OpenLAM dataset collection (refer to Supporting Information \cref{tab:si-datasets}). 
This collection encompasses a wide array of chemical and material systems, including inorganic materials, catalysis, small molecules, biomolecules, and reactions, all under diverse DFT settings. 
Detailed hyperparameter information for the models can be found in Supporting Information \cref{tab:si-model}.
Unless specified otherwise, the MPtrj task head is employed by default across all benchmarks.


\begin{table}
    \centering
    \caption{
    Summary of LAMs benchmarked in this study.
    The table includes the model name, number of parameters, training dataset, and cutoff radius. 
    For conservative models that calculate force as the negative gradient of energy, ``Direct Force Prediction'' is indicated as ``No''.
    }
    {\fontsize{9}{9.5}\selectfont
    \begin{tabular}{lcccccc}
    \toprule
\textbf{Model} & \textbf{\# Parameters} & \textbf{Training Set} & \textbf{\makecell{Direct Force\\Prediction}} & \textbf{Cutoff Radius (\r{A})}\\
        \midrule
DPA-3.1-3M\cite{zhang2025graphneuralnetworkera} & 3.27M & OpenLAM & No & 6.0\\
DPA-2.4-7M\cite{zhang2024dpa} & 6.64M & OpenLAM & No & 6.0\\
MACE-MP-0 medium\cite{batatia2023foundation} & 4.69M & MPtrj & No & 6.0\\
MACE-MPA-0 medium\cite{batatia2023foundation} & 9.06M & MPtrj, sAlex & No & 6.0\\
Orb-v2\cite{neumann2024orb} & 25.2M & MPtrj, Alex & Yes & 10.0 \\
Orb-v3-conservative-inf-mpa\cite{rhodes2025orbv3atomisticsimulationscale} & 25.5M & MPtrj, Alex & No & 6.0 \\
SevenNet-l3i5\cite{park2024scalable} & 1.17M & MPtrj & No & 5.0\\
SevenNet-MF-ompa\cite{kim_sevennet_mf_2024} & 25.7M & OMat24, MPtrj, sAlex & No &6.0\\
MatterSim-v1-5M\cite{yang2024mattersimdeeplearningatomistic} & 4.55M & MattterSim & No &5.0\\
GRACE-2L-OAM\cite{PhysRevX.14.021036}& 12.6M & MPtrj & No & 6.0\\

        \bottomrule
\end{tabular}}
\label{tab:models}
\end{table}

\subsection{LAMBench Implementation}
\label{sec:implementation}

Benchmarking LAMs involves repeatedly performing computations using various combinations of models and test tasks, followed by aggregating and visualizing the benchmarking results.
The combination of models and tasks generates a substantial job array, rendering manual submission inefficient. 
To automate and enhance the benchmarking process, we developed the LAMBench-toolkit, as illustrated in \cref{fig:toolkit}.
By offering a structured and automated benchmarking framework, LAMBench significantly facilitates the comprehensive evaluation and comparison of LAMs. 
The LAMBench-toolkit is openly available under the MIT License at \url{github.com/deepmodeling/lambench}.
An interactive leaderboard is provided at \url{https://www.aissquare.com/openlam?tab=Benchmark}.

In the LAMBench-toolkit, model definitions, test tasks, and workflow management are implemented as distinct modules. 
This modular design allows LAM developers to effortlessly incorporate new models and test cases into the toolkit, while also facilitating the efficient maintenance of existing components.

\paragraph{Models.} 
Each LAM within the LAMBench-toolkit is specified via a configuration file that includes the associated Python package name and the path for loading model weights.
Models engage with test tasks through the Atomic Simulation Environment (ASE) calculator interface\cite{larsen2017atomic}, offering a standardized approach for model-task interaction. 
Developers can seamlessly integrate new LAMs into LAMBench-toolkit by providing their ASE calculators.

\paragraph{Tasks.}
The tasks module implements the benchmark tasks that researchers intend to perform on LAMs.
Each task explicitly delineates the calculation workflow to evaluate a specific capability of a model and provides output metrics, such as the error in calculating a particular value, to quantify performance.
Upon completion of a task, the resulting metrics and model information are uploaded to a database, facilitating easier data analysis and management. 
Additionally, the database is utilized to identify and skip duplicate computational jobs at the start of each task.

\paragraph{Workflow.} 
Within the context of benchmarking LAMs, a computational workflow denotes a structured sequence of computational steps aimed at assessing model performance across diverse models and datasets. 
The workflow module of LAMBench orchestrates these benchmarking steps, efficiently handling job submissions, executions, and the subsequent aggregation and analysis of results. 
This design ensures that developers of models and test tasks are not burdened by the specifics of execution on computational resources, nor are they required to manually collect and analyze the results.

The array of computational steps generated by the workflow module is submitted to computational resources through Dflow\cite{liu2024dflow}, a functional programming interface designed for scientific computing workflows.
In our experiments, jobs are executed on cloud instances equipped with an NVIDIA V100 32GB GPU via the Bohrium Cloud Platform. 
Various other computational resources, including high-performance supercomputers and local hardware, are also supported.
Upon job completion, the workflow retrieves results from the database, calculates the metrics, and updates the visualization plots on the webpage frontend. 
This process enables researchers to intuitively analyze and interpret the benchmarking outcomes.

\subsection{Generalizability metrics }
\label{sec:generalizability-score}

In assessing the generalizability of models, the primary metrics employed are the mean absolute error (MAE) and root mean square error (RMSE) for specific predictions across test sets within various domains. 
Direct comparison of model performance using these metrics can be challenging due to the extensive number of metrics generated for each model. 
This complexity arises because each model typically provides numerous error measurements across different prediction types and test sets. 
Thus, a more integrated approach to evaluation is required to effectively compare the generalizability of different models, considering the multitude of error metrics involved.

In this study, we denote the error metric as \( M^m_{k,p,i} \), where \( m \) indicates the model, \( k \) denotes the domain index, \( p \) signifies the prediction index, and \( i \) represents the test set index.
For instance, in force field tasks, the domains include \domainsm, \domainim, and \domaincat, such that \( k \in \{\text{Molecules, Inorganic Materials, Catalysis}\} \). The prediction types are categorized as energy (\(E\)), force (\(F\)), or virial (\(V\)), with \( p \in \{E, F, V\} \).
For the specific domain of \domainsm, the test sets are indexed as $ i \in \{\text{MD22, ANI-1x, AIMD-Chig}\}$.
To facilitate a fair comparison, the error metric is truncated and normalized against the error metric of a baseline model (dummy model) as follows:
\begin{equation} \label{eq:norm-e}
    \hat{M}^m_{k,p,i} = \min\left(\frac{M^m_{k,p,i}}{M^{\mathrm{dummy}}_{k,p,i}}, 1\right)
\end{equation}
This baseline model predicts energy based solely on the chemical formula, disregarding any structural details, thereby providing a reference point for evaluating the improvement offered by more sophisticated models.
For each domain, we compute the log-average of normalized metrics across all datasets  within this domain by
\begin{equation} 
    \bar{M}^m_{k,p} = \exp\left(\frac{1}{n_{k,p}}\sum_{i=1}^{n_{k,p}}\ln \hat{M}^m_{k,p,i}\right),
\end{equation}
where $n_{k,p}$ denotes the number of test sets for domain $k$ and prediction type $p$.
Subsequently, we calculate a weighted dimensionless domain error metric to encapsulate the overall error across various prediction types:
\begin{equation}
    \bar{M}^m_{k}  = \sum_p w_{p} \bar{M}^m_{k,p} \Bigg/ \sum_p w_{p},
\end{equation}
where \( w_{p} \) denotes the weights assigned to each prediction type \(p\).

Finally the overall generalizability error metric of a model across all the domains is defined by the average of the domain error metrics,
\begin{equation}
{\bar{M}^m}= \frac{1}{n_D}\sum_{k=1}^{n_D}{\bar{M}^m_{k}}, 
\end{equation}
where $n_D$ denotes the number of domains under consideration. 
The dimensionless error metric ${\bar{M}^m}$ allows for the comparison of generalizability across different models.
It reflects the overall generalization capability across all domains, prediction types, and test sets, with a lower value indicating superior performance. 
The only tunable parameter is the weights assigned to prediction types, thereby minimizing arbitrariness in the comparison system.

For the force field generalizability tasks, we adopt RMSE as the primary error metric.
The prediction types include energy and force, with weights assigned as \( w_E = w_F = 0.5 \). 
When periodic boundary conditions are assumed and virial labels are available, virial predictions are also considered. 
In this scenario, the prediction weights are adjusted to \( w_E = w_F = 0.45 \) and \( w_V = 0.1 \).
The resulting error metric after averaging over all domains is referred to as \(\bar{M}^{m}_{\mathrm{FF}}\).

For the domain-specific property calculation tasks, we adopt the MAE as the primary error metric.
In the \domainim\ domain, the MDR phonon benchmark predicts maximum phonon
frequency, entropy, free energy, and heat capacity at constant volume, while the elasticity benchmark evaluates the shear and bulk moduli. 
Each prediction type is assigned an equal weight of 1/6.
In the \domainsm\ domain, the TorsionNet500 benchmark evaluates the torsion profile energy, torsional barrier height, and the number of molecules for which the predicted torsional barrier height error exceeds 1 kcal/mol. 
The Wiggle150 benchmark assesses the relative conformer energy profile. 
Each prediction type in this domain is assigned a weight of 1/4. 
In the \domaincat\ domain, the OC20NEB-OOD benchmark evaluates the energy barrier, reaction energy change (delta energy), and the percentage of reactions with predicted energy barrier errors exceeding 0.1 eV for three reaction types: transfer, dissociation, and desorption. Each prediction type in this domain is assigned a weight of 1/5.
The resulting error metric after averaging over all domains is denoted as \(\bar{M}^{m}_{\mathrm{PC}}\).

\subsection{Applicability metrics}
\label{sec:applicability-score}

The applicability metrics incorporate both efficiency and stability and are computed differently from the generalizability metrics due to the absence of a dummy model baseline.
To evaluate efficiency, we define an efficiency metric,  \(M_E^m\) , by normalizing the average inference time (with unit \(\mathrm{\mu s/atom}\)), $\bar \eta^m$, of a given LAM measured over 900 configurations with respect to an artificial reference value, thereby rescaling it to a range between zero and positive infinity. A larger value indicates the higher efficiency.
\begin{equation}
M_{\mathrm{E}}^m = \frac{\eta^0 }{\bar \eta^m },\quad \eta^0= 100\  \mathrm{\mu s/atom}, \quad \bar \eta^m = \frac{1}{900}\sum_{i}^{900} \eta_{i}^{m},
\end{equation}
where $\eta_{i}^{m}$ is the inference time of configuration $i$ for model $m$.

Stability is quantified by measuring the total energy drift in NVE simulations across nine structures.
For each simulation trajectory, an instability metric is defined based on the magnitude of the slope obtained via linear regression of total energy per atom versus simulation time. 
A tolerance value, $5\times10^{-4} \ \mathrm{eV/atom/ps}$,  is determined as three times the statistical uncertainty in calculating the slope from a 10 ps NVE-MD trajectory using the MACE-MPA-0 model.
If the measured slope is smaller than the tolerance value, the energy drift is considered negligible. 
We define the dimensionless measure of instability for structure $i$ as follows:
\begin{equation}
M^m_{\mathrm{IS},i} = 
\begin{cases}
\max\left(0, \log_{10}({\Phi_{i}}/{\Phi_{\mathrm{tol}}})\right), & \text{if success} \\
5, & \text{otherwise}
\end{cases}, \quad
\Phi_{\mathrm{tol}} = 5\times10^{-4} \ \mathrm{eV/atom/ps},
\end{equation}
where $\Phi_i$ represents the total energy drift, and $\Phi_{\mathrm{tol}}$ denotes the tolerance.
This metric indicates the relative order of magnitude of the slope compared to the tolerance. 
In cases where a MD simulation fails, a penalty of 5 is assigned, representing a drift five orders of magnitude larger than the typical statistical uncertainty in measuring the slope.
The final instability metric is computed as the average over all nine structures.
\begin{equation}
M^m_{\mathrm{IS}} = \frac{1}{9}\sum_{i=1}^{9} M^m_{\mathrm{IS},i}
\end{equation}
This result is bounded within the range $[0, +\infty)$, where a lower value signifies greater stability.

\section*{Author contributions}
A.P., C.C., L.Z., and H.W. conceived the idea of this work and developed the benchmark workflow. 
M.G., D.Z., and C.Z. collected and analyzed the results across the benchmark tasks. 
W.J. and Y.W. performed DFT calculations for dataset relabeling. 
A.L. provided guidance on the adaptation of the phonon benchmark. 
C.W. and W.E. provided guidance on the workflow design. 
All authors contributed to the discussions and revised the manuscript.

\section*{Acknowledgments}

The computational resources utilized in this work were provided by the AI for Science Institute.
The work is supported by National Key Research and Development Program of China Grant No.~2022YFA1004300.
A.L. acknowledges funding from the Horizon Europe MSCA Doctoral network grant n.101073486, EUSpecLab, funded by the European Union.


\renewcommand{\thefigure}{S-\arabic{figure}}
\setcounter{figure}{0}
\renewcommand{\thetable}{S-\arabic{table}}
\setcounter{table}{0}
\renewcommand{\thesection}{S-\arabic{section}}
\setcounter{section}{0}







\makeatletter
\def\maketitle{
\@author@finish
\title@column\titleblock@produce
\suppressfloats[t]}
\makeatother

\title{Supporting Information}
\maketitle
\section{DPA Model Configurations}
\begin{table}[h]
    \caption{Hyperparameters for DPA-2.4-7M model}
    \centerline{
    \small{
    \begin{tabular}{llc}
        \toprule
        \textbf{Module} & \textbf{Hyperparameter} & \textbf{DPA-2.4-7M} \\
        \midrule
        \multirow{3}{*}{Repinit} 
        & nsel & 120 \\
        & neuron & 3$\times$64 \\
        & three\_body\_neuron & 3$\times$32 \\
        \midrule
        \multirow{4}{*}{Repformer} 
        & nsel & 40 \\
        & nlayers & 6 \\
        & g1\_dim & 384 \\
        & g2\_dim & 96 \\
        \midrule
        Activation & & tanh \\
        \midrule
        \multirow{6}{*}{Training}
        & batch size & ``auto:256" \\
        & nGPUs & 120 $\times$ A800-40GB \\
        & start lr & 1e-3 \\
        & stop lr & 1e-5 \\
        & precision & float32 \\
        & steps trained & 8M \\
        \bottomrule
    \end{tabular}
    }
    }
    \label{tab:si-model}
\end{table}
\clearpage

\newcolumntype{L}[1]{>{\scriptsize\raggedright\arraybackslash}p{#1}}
\begin{longtable}{L{4cm} L{2.5cm} L{2cm} L{3cm} L{3cm} L{1cm}}
    \caption{Summary of datasets, including the dataset name, number of frames, average number of atoms, data cleaning procedures, DFT level, and dataset weight used in training.} \label{tab:si-datasets} \\
    \toprule
    \textbf{Dataset Name} & \textbf{\# Frames} & \textbf{Avg. \# atoms} & \textbf{Data Cleaning} & \textbf{Level of Theory} & \textbf{Weight} \\
    \midrule
    \endfirsthead
    
    \multicolumn{6}{c}{{\tablename\ \thetable{} -- Continued from previous page}} \\
    \toprule
    \textbf{Dataset Name} & \textbf{\# Frames} & \textbf{Avg. \# atoms} & \textbf{Data Cleaning} & \textbf{DFT Level} & \textbf{Weight} \\
    \midrule
    \endhead
    
    \bottomrule
    \multicolumn{6}{r}{{Continued on next page}} \\
    \endfoot
    
    \bottomrule
    \endlastfoot

    Alex2D\cite{wang2023symmetry} & \makecell[l]{trn: 1,223,831\\ val: 135,810} & 9.9 & Exclude frames with energies $>$ 0~eV/atom, maximum absolute force $>$ 5~eV/\AA, maximum absolute virial/atom $>$ 8~eV/atom; Trajectory first/last 3 frames + every 10 frame & PBE/PAW, 520~eV, 0.4 \AA$^{-1}$ & 1.83 \\
    \addlinespace
    MPTraj\cite{Deng2023} & \makecell[l]{trn: 1,401,956\\ val: 110,918} & 31.35 & According to the distribution difference between MACE-MP-0 prediction and label, the union set of top 0.1\% with the largest uncorrected energy difference and the top 0.1\% with the largest force difference were excluded, resulting in a total of 3123 frames excluded. & PBE (+U)/PAW, 520~eV, 0.04 \AA$^{-1}$ & 4.66\\
    \addlinespace
    OC20M\cite{chanussot2021open} & \makecell[l]{trn: 20,000,000\\ val: 999,866} & 73.1 & - & rPBE/PAW, 350eV & 11.65\\
    \addlinespace
    OC22\cite{doi:10.1021/acscatal.2c05426} & \makecell[l]{trn: 8,194,770\\ val: 394,727} & 79.8 & Exclude frames with energies $>$ 0eV/atom, maximum absolute force $>$ 10eV/\AA & PBE (+U)/PAW, 500~eV & 8.14\\
    \addlinespace    
    ODAC23\cite{sriram2024open} & \makecell[l]{trn: 2,682,332\\ val: 63,623} & 203 & Exclude frames with energy/atom $>$ 0.5~eV/atom or $<$ -0.2~eV/atom, and maximum absolute force $>$ 25~eV/\AA; Trajectory first/last 3 frames + every 20th frame & PBE-D3/PAW, 600~eV, gamma & 1.19\\
    \addlinespace
    OMAT24\cite{barrosoluque2024openmaterials2024omat24} & \makecell[l]{trn: 100,568,820\\ val: 1,074,647} & 18.6 & Exclude frames with energies $>$ 0~eV/atom, $<$ -25eV/atom, maximum absolute force $>$ 50eV/\AA & PBE (+U)/PAW & 22.16\\
    \addlinespace
    SPICE2\cite{eastman2024nutmeg} & \makecell[l]{trn: 1,621,168\\ val: 180,089} & 35.6 & Exclude frames with energies $<$ -10000~eV/atom, maximum absolute force $>$ 15eV/\AA & $\omega$B97M-D3(BJ)/def2-TZVPPD & 5.39\\
    \addlinespace
    Transition1x\cite{schreiner2022transition1x} & \makecell[l]{trn: 7,632,328\\ val: 967,454} & 13.91 & - & $\omega$B97X/6-31G(d) & 2.28 \\
    \addlinespace
    UniPero\cite{wu2023universal} & \makecell[l]{trn: 14,487\\ val: 1,357} & 53.64 & - & PBEsol/LCAO, 1360~eV, 0.189 \AA$^{-1}$ & 0.77\\
    \addlinespace
    Dai2023Alloy\cite{Dai2023Alloy} & \makecell[l]{trn: 71,482\\ val: 1,240} & 20.99 & - & PBE/Norm-conserving (NC), 1360~eV, 0.094 \AA$^{-1}$ & 0.67\\
    \addlinespace
    Zhang2023Cathode\cite{Zhang2023Cathode} & \makecell[l]{trn: 88,692\\ val: 9,695} & 46.28 & - & PBE (+U)/PAW, 520~eV, 0.25 \AA$^{-1}$ & 1.64\\
    \addlinespace
    Gong2023Cluster\cite{Gong2023Cluster} & \makecell[l]{trn: 143,418\\ val: 15,331} & 23.67 & - & PBE-D3/TZV2P, 400-1000 Ry & 1.07 \\
    \addlinespace
    Li2025APEX\cite{li2025apex} & \makecell[l]{trn: 24,097\\ val: 100} & 24.03 & - & PBE/NC, 1360~eV, 0.15 \AA$^{-1}$ & 0.42\\
    \addlinespace
    Shi2024Electrolyte\cite{Shi2024Electrolyte} & \makecell[l]{trn: 65,393\\ val: 3,438} & 192.69 & - & PBE-D3, 800 Ry & 5.85\\
    \addlinespace
    Shi2024SSE\cite{Shi2024SSE} & \makecell[l]{trn: 125,083\\ val: 6,587} & 48.33 & - & PBE-sol/LCAO, 1360~eV, 0.28 \AA$^{-1}$ & 2.03\\
    \addlinespace
    Yang2023ab\cite{yang2024ab} & \makecell[l]{trn: 1,379,956\\ val: 24,257} & 33.47 & - & $\omega$B97X-D/6-31G** & 4.67\\
    \addlinespace
    Li2025General\cite{Li2025General} & \makecell[l]{trn: 14,024,587\\ val: 1,558,260} & 17.8 & Exclude maximum absolute force $>$ 20eV/\AA & GFN2-xTB & 7.95\\
    \addlinespace
    Huang2021Deep-PBE\cite{huang2021deep} & \makecell[l]{trn: 17,582\\ val: 886} & 218.59 & - & PBE/PAW, 650~eV, 0.26 \AA$^{-1}$ & 1.72\\
    \addlinespace
    Liu2024Machine\cite{liu2024machine} & \makecell[l]{trn: 215,481\\ val: 23,343} & 70.36 & - & PBE/LCAO, 1360~eV, 0.151 \AA$^{-1}$ & 3.88 \\
    \addlinespace
    Zhang2021Phase\cite{zhang2021phase} & \makecell[l]{trn: 46,077\\ val: 2,342} & 172.3 & - & SCAN/PAW, 1500~eV, 0.5 \AA$^{-1}$ & 4.39\\
    \addlinespace
    Jiang2021Accurate\cite{jiang2021accurate} & \makecell[l]{trn: 138,194\\ val: 3,965} & 23.16 & - & PBE/PAW, 650~eV, 0.1 \AA$^{-1}$ & 0.51\\
    \addlinespace
    Chen2023Modeling\cite{chen2023modeling} & \makecell[l]{trn: 6,449\\ val: 276} & 16.94 & - & SCAN/PAW, 650~eV, 0.1 \AA$^{-1}$ & 0.03\\
    \addlinespace
    Unke2019PhysNet\cite{unke2019physnet} & \makecell[l]{trn: 2,594,609\\ val: 136,571} & 21.38 & - & revPBE-D3(BJ)/def2-TZVP & 4.09\\
    \addlinespace
    Wen2021Specialising\cite{wen2021specialising} & \makecell[l]{trn: 10,054\\ val: 474} & 20.02 & - & PBE/PAW, 650~eV, 0.1 \AA$^{-1}$ & 0.05 \\
    \addlinespace
    Wang2022Classical\cite{wang2022classical} & \makecell[l]{trn: 14,935\\ val: 738} & 24.47 & - & PBE/PAW, 650~eV, 0.1 \AA$^{-1}$ & 0.07\\
    \addlinespace
    Wang2022Tungsten\cite{wang2022tungsten} & \makecell[l]{trn: 42,297\\ val: 2,100} & 24.91 & - & PBE/PAW, 600~eV, 0.16 \AA$^{-1}$ & 0.12 \\
    \addlinespace
    Wu2021Deep\cite{wu2021deep} & \makecell[l]{trn: 27,660\\ val: 917} & 96 & - & PBE/PAW, 600~eV, 2$\times$2$\times$2 kpt & 0.95\\
    \addlinespace
    Huang2021Deep-PBEsol\cite{huang2021deep} & \makecell[l]{trn: 7,502\\ val: 384} & 160.8 & - & PBE-sol/PAW, 650~eV, 0.26 \AA$^{-1}$ & 0.83\\
    \addlinespace
    Wang2021Generalizable\cite{wang2021generalizable} & \makecell[l]{trn: 64,239\\ val: 2,256} & 21.85 & - & PBE-D3/PAW, 650~eV, 0.1 \AA$^{-1}$ & 0.26\\
    \addlinespace
    Wu2021Accurate\cite{wu2021accurate} & \makecell[l]{trn: 11,621\\ val: 568} & 21.73 & - & PBE/PAW, 700~eV &0.14 \\
    \addlinespace
    Tuo2023Hybrid\cite{tuo2023spontaneous} & \makecell[l]{trn: 48,078\\ val: 2,530} & 45.11 & - & PBE-D3(BJ)/PAW, 500~eV, 0.16 \AA$^{-1}$ & 0.59\\
    \bottomrule
\end{longtable}

\clearpage
\section{Original Error Metrics for Generalizability Domain Specific Tasks}
\begin{table}[h]
    \centering
    \caption{Phonon related property prediction on the MDR phonon benchmark.}
    {\small
    \begin{tabular}{lcccc}
        \toprule
        \textbf{Model} & \textbf{\makecell{MAE $\omega_{\text{max}}$\\ (K)}} & \textbf{\makecell{MAE $S$\\ (J/K/mol)}} & \textbf{\makecell{MAE $F$ \\ (kJ/mol)}} & \textbf{\makecell{MAE $C_V$ \\(J/K/mol)}} \\
        \midrule
        DPA-3.1-3M-DomainMatch & 32.4 & 24.4 & 7.0 & 9.1\\
        DPA-3.1-3M & 38.4 & 33.4 & 9.9 & 10.9\\
        DPA-2.4-7M & 58.9 & 45.6 & 18.1 & 12.8\\
        MACE-MP-0 & 61.0 & 59.6 & 23.8 & 13.1 \\
        MACE-MPA-0 & 29.7 & 19.8 & 7.9 & 5.8 \\
        Orb-v2 & 308.0 & 446.5 & 184.3 & 58.5 \\
        Orb-v3 & 72.1 & 31.5 & 7.3 & 13.6 \\
        SevenNet-l3i5 & 25.6 & 25.9 & 9.6 & 4.9\\
        SevenNet-MF-ompa & 14.9 & 10.5 & 4.1 & 3.1\\
        MatterSim-v1-5M & 16.4 & 15.2 & 5.2 & 3.1\\
        GRACE-2L-OAM & 19.5 & 14.1 & 5.5 & 3.7\\
        dummy & 1188.3 & 764.8 & 125.1 & 547.4 \\
        \bottomrule

    \end{tabular}
    }
    \label{tab:si-phonon}
\end{table}

\begin{table}[h]
    \centering
    \caption{Benchmark results on elasticity related property prediction.}
    {\small
    \begin{tabular}{lcc}
        \toprule
        \textbf{Model} & \textbf{\makecell{MAE $_{GVRH}$\\}} & \textbf{\makecell{MAE $_{KVRH}$\\}}  \\
        \midrule
        DPA-3.1-3M-DomainMatch & 9.83 & 8.68 \\
        DPA-3.1-3M & 10.77 & 10.13\\
        DPA-2.4-7M & 17.76 & 16.46\\
        MACE-MP-0 & 26.20 & 11.01 \\
        MACE-MPA-0 & 10.27 & 15.03 \\
        Orb-v2 & 66.07 & 44.08 \\
        Orb-v3 & 9.75 & 7.58 \\
        SevenNet-l3i5 & 19.42 & 9.93\\
        SevenNet-MF-ompa & 9.54 & 9.46\\
        MatterSim-v1-5M & 12.75 & 14.95\\
        GRACE-2L-OAM & 9.14 & 7.46\\
        dummy & 67.54 & 136.26 \\
        \bottomrule

    \end{tabular}
    }
    \label{tab:si-elasticity}
\end{table}

\begin{table}[h]
    \centering
    \caption{Torsional MAE and MAEB errors between LAM predictions and its reference DFT labels on the TorsionNet-500 Benchmark.}
    {\small
    \begin{tabular}{lccc}
        \toprule 
        \textbf{Model} & \textbf{\makecell{MAE \\ (kcal/mol)}} & \textbf{\makecell{MAEB\footnotemark[1]\\ (kcal/mol)}} & \textbf{$\text{NABH}_{h}$\footnotemark[2]} \\
        \midrule
        DPA-3.1-3M-DomainMatch & 0.338 & 0.551 & 76 \\
        DPA-3.1-3M & 0.825 & 1.382 & 298 \\
        DPA-2.4-7M & 0.739 & 1.229 & 265 \\
        MACE-MP-0 & 1.652 & 2.437 & 356  \\
        MACE-MPA-0 & 1.468 & 2.150 & 339  \\
        Orb-v2 & 1.248 & 2.109 & 346 \\
        Orb-v3 & 0.988 & 1.421 & 268 \\
        SevenNet-l3i5 & 1.127 & 1.641 & 308 \\
        SevenNet-MF-ompa & 1.619 & 2.338  & 360\\
        MatterSim-v1-5M & 1.360 & 2.234 & 342 \\
        GRACE-2L-OAM & 1.394 & 1.985 & 309 \\
        dummy & 2.494 & 5.818 & 493 \\
        \bottomrule
        \\
    \end{tabular}
    }
    \label{tab:si-torsion500}

\footnotemark[1]{\footnotesize The mean-absolute-error of the torsional barrier height, defined as the difference between the minimum and the maximum energy points during the torsional rotation.}
\footnotemark[2]{\footnotesize The number of molecules (total: $N_{\text{mols}}$ = 500) for which the model prediction of torsional barrier height has an error of more than 1 kcal/mol.}
\end{table}

\begin{table}[h]
    \centering
    \caption{Prediction of conformer relative energies on the Wiggle150 benchmark.}
    {\small
    \begin{tabular}{lc}
        \toprule
        \textbf{Model} & \textbf{\makecell{MAE \\ (kcal/mol)}} \\
        \midrule
        DPA-3.1-3M-DomainMatch & 1.72 \\
        DPA-3.1-3M & 5.67\\
        DPA-2.4-7M & 14.84\\
        MACE-MP-0 & 26.60 \\
        MACE-MPA-0 & 14.91 \\
        Orb-v2 & 6.46 \\
        Orb-v3 & 11.92 \\
        SevenNet-l3i5 & 13.88\\
        SevenNet-MF-ompa & 10.97\\
        MatterSim-v1-5M & 10.73\\
        GRACE-2L-OAM & 12.14\\
        dummy & 105.06 \\
        \bottomrule

    \end{tabular}
    }
    \label{tab:si-wiggle}
\end{table}

\begin{table}[h]
    \centering
    \caption{Energy barrier prediction on the OC20NEB-OOD Benchmark.\protect\footnotemark[1]
    } 
    {\small
    \begin{tabular}{lccccc}
        \toprule 
        \textbf{Model} 
        & \textbf{\makecell{MAE$_{Ea}$\footnotemark[2] \\ (eV)}} 
        & \textbf{\makecell{MAE$_{dE}$\footnotemark[3] \\ (eV)}} 
        & \textbf{$\phi_\text{Transfer}$\footnotemark[4]} 
        & \textbf{$\phi_\text{Dissociation}$\footnotemark[4]} 
        & \textbf{$\phi_\text{Desorption}$\footnotemark[4]} \\
        \midrule
        DPA-3.1-3M-DomainMatch & 1.172 & 0.158 & 65.1 & 69.0 & 59.8 \\
        DPA-3.1-3M      & 1.203 & 0.234 & 66.9 & 63.9 &  81.1\\
        DPA-2.4-7M      & 1.271 & 0.306 & 68.6 & 77.9 & 69.3 \\
        MACE-MP-0       & 1.468 & 0.580 & 76.6 & 84.8 & 90.6 \\
        MACE-MPA-0      & 1.459 & 0.566 & 72.6 & 84.8 & 94.5 \\
        Orb-v2          & 2.682 & 1.729 & 66.3 & 74.1 & 78.7 \\
        Orb-v3          & 2.298 & 1.470 & 61.7 & 72.2 & 87.4 \\
        SevenNet-l3i5   & 1.423 & 0.546 & 77.7 & 76.0 & 87.4 \\
        SevenNet-MF-ompa& 2.070 & 1.278 & 66.9 & 68.4 & 92.9 \\
        MatterSim-v1-5M & 2.686 & 2.018 & 76.0 & 84.8 & 83.5 \\
        GRACE-2L-OAM    & 1.583 & 0.704 & 65.1 & 72.2 & 90.6 \\
        dummy           & 2.360 & 0.950 & 95.4 & 99.4 & 95.3 \\
        \bottomrule
    \end{tabular}
    }
    \label{tab:si-neb}

    \footnotemark[1]{\footnotesize Only 460 out-of-distribution trajectories were used in this task.}
    \footnotemark[2]{\footnotesize The mean absolute error of the reaction barrier.}
    \footnotemark[3]{\footnotesize The mean absolute error of the reaction energy.}
    \footnotemark[4]{\footnotesize Percentage of reactions with an energy barrier error exceeding 0.1 eV. The analysis considers three reaction types: transfer, dissociation, and desorption.}
\end{table}

\clearpage

\section{Structures for Conservativeness Benchmark}
\begin{figure}[ht]
    \centering
    \includegraphics[width=0.8\linewidth]{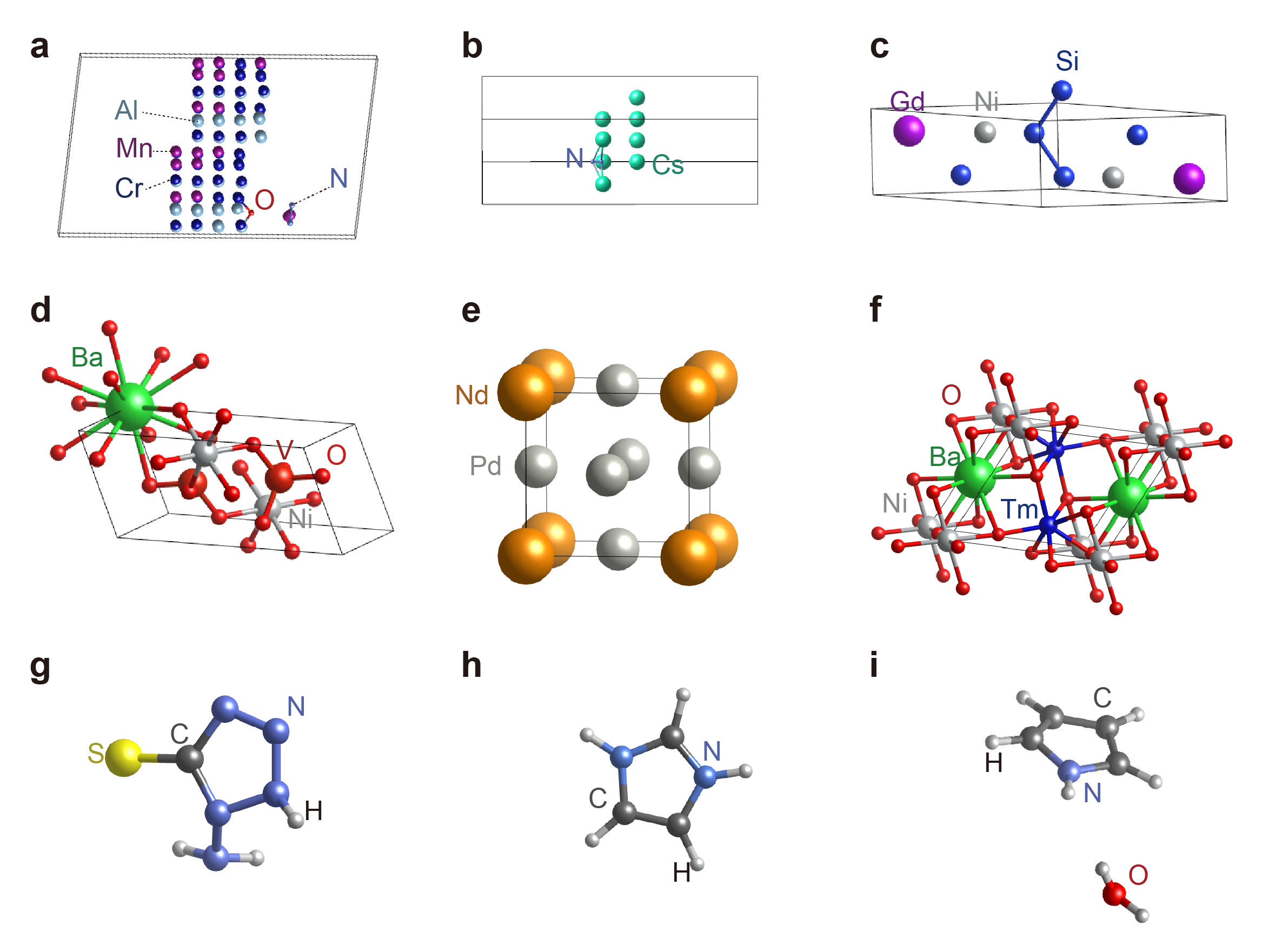}
    \caption{Structures used for the Conservativeness Benchmark. 
        (a) H\allowbreak$_{2}$Al\allowbreak$_{32}$Cr\allowbreak$_{48}$Mn\allowbreak$_{16}$N\allowbreak$_{2}$O; 
        (b) Cs\allowbreak$_{8}$N\allowbreak$_{2}$; 
        (c) Gd\allowbreak$_{2}$Ni\allowbreak$_{2}$Si\allowbreak$_{4}$; 
        (d) NdPd\allowbreak$_{3}$; 
        (e) BaNi\allowbreak$_{2}$O\allowbreak$_{8}$V\allowbreak$_{2}$; 
        (f) BaNiO\allowbreak$_{5}$Tm\allowbreak$_{2}$; 
        (g) CH\allowbreak$_{3}$N\allowbreak$_{5}$S; 
        (h) C\allowbreak$_{3}$H\allowbreak$_{5}$N\allowbreak$_{2}$; 
        (i) C\allowbreak$_{4}$H\allowbreak$_{7}$NO.}
    \label{fig:si-nve_struct}
\end{figure}
\clearpage

\section{Inference Efficiency Convergence Test}
\begin{figure}[h]
    \centering
    \includegraphics[width=0.8\linewidth]{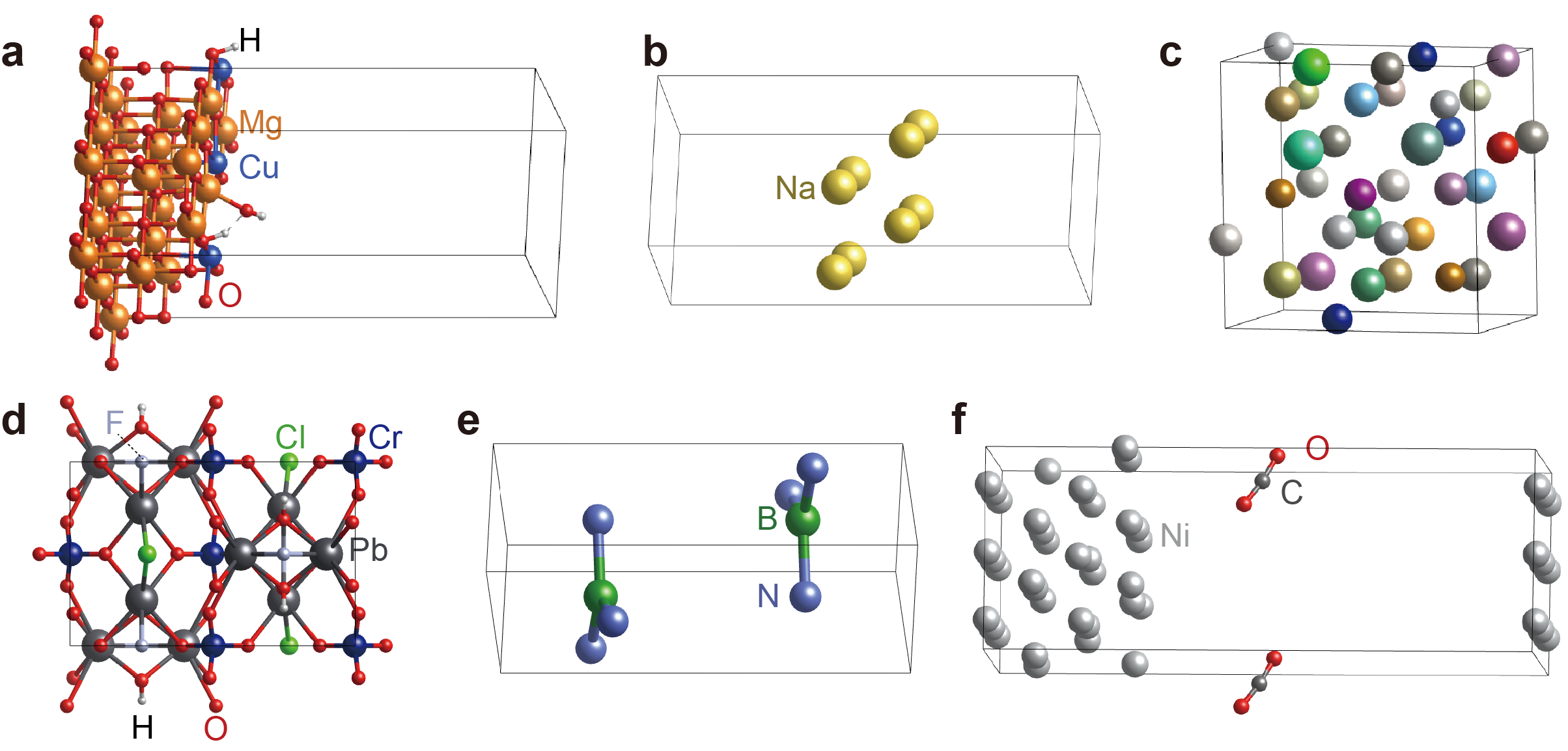}
    \caption{The structures in unit cell for convergence test in efficiency benchmark.
        (a) H\allowbreak Cu\allowbreak Mg\allowbreak$_{11}$\allowbreak O\allowbreak$_{12}$; 
        (b) Na; 
        (c) High-entropy alloy: Ag\allowbreak$_{2}$\allowbreak Au\allowbreak Co\allowbreak Cr\allowbreak Cu\allowbreak Fe\allowbreak$_{2}$\allowbreak Hf\allowbreak Ir\allowbreak Lu\allowbreak Mn\allowbreak Mo\allowbreak$_{2}$\allowbreak Nb\allowbreak$_{2}$\allowbreak Ni\allowbreak$_{2}$\allowbreak Pd\allowbreak Pt\allowbreak$_{2}$\allowbreak Rh\allowbreak Ru\allowbreak Sc\allowbreak$_{2}$\allowbreak Ta\allowbreak$_{2}$\allowbreak Ti\allowbreak$_{2}$\allowbreak V\allowbreak W\allowbreak$_{3}$\allowbreak Y\allowbreak Zn\allowbreak Zr; 
        (d) H\allowbreak$_{2}$\allowbreak Cl\allowbreak Cr\allowbreak$_{2}$\allowbreak F\allowbreak O\allowbreak$_{10}$\allowbreak Pb\allowbreak$_{4}$; 
        (e) BN; 
        (f) C\allowbreak Ni\allowbreak$_{36}$\allowbreak O\allowbreak$_{2}$.}
    \label{fig:si-eff_struc}
\end{figure}

\begin{figure}[h]
    \centering
    \includegraphics[width=0.8\linewidth]{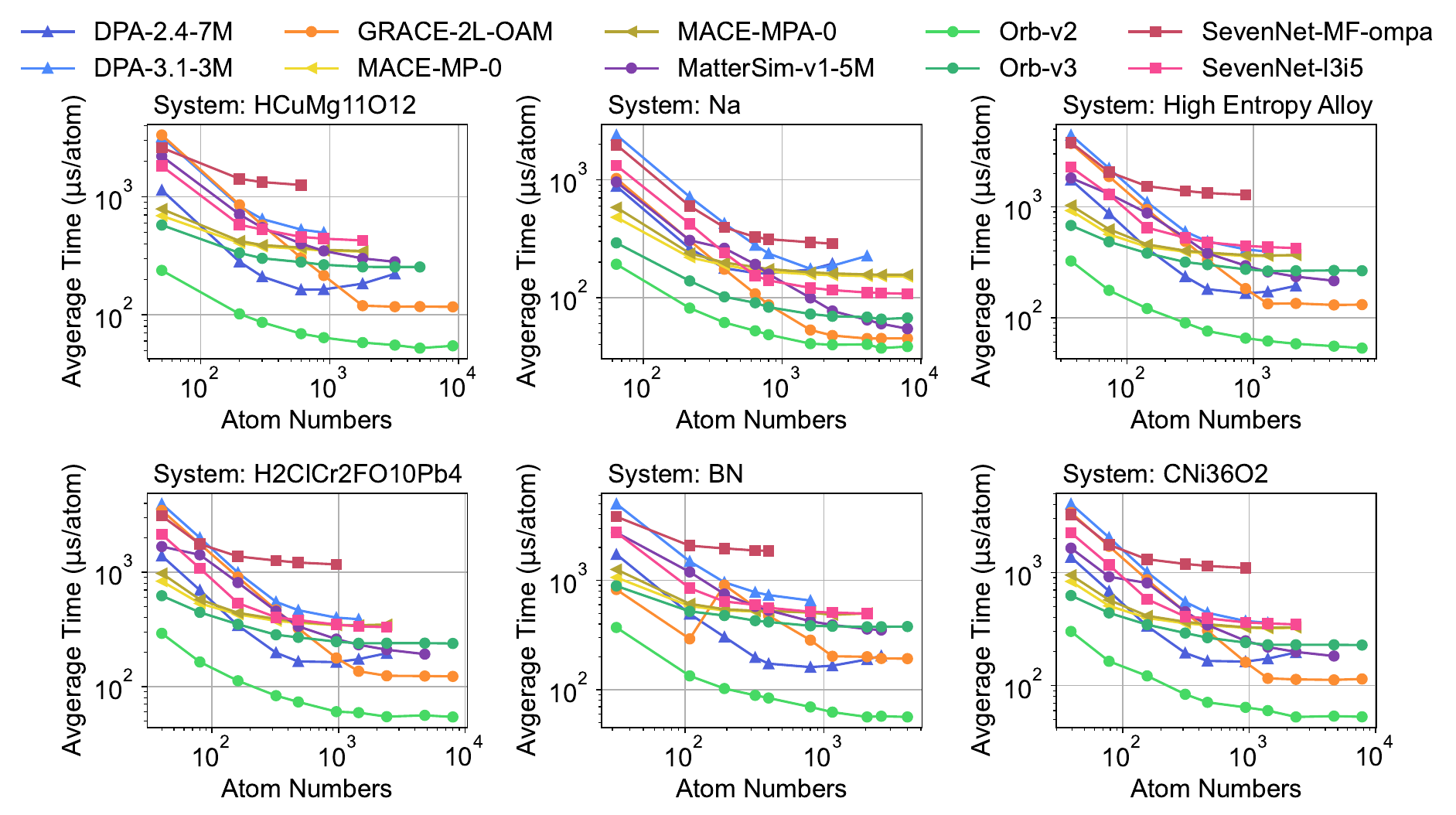}
    \caption{The convergence test of efficiency benchmark with respect to atom numbers.}
    \label{fig:si-eff_converge}
\end{figure}
\clearpage

\section{The Coordination Number Analysis of Configurations in Efficiency Benchmarks}

\begin{figure}[h]
    \centering
    \includegraphics[width=1.0\linewidth]{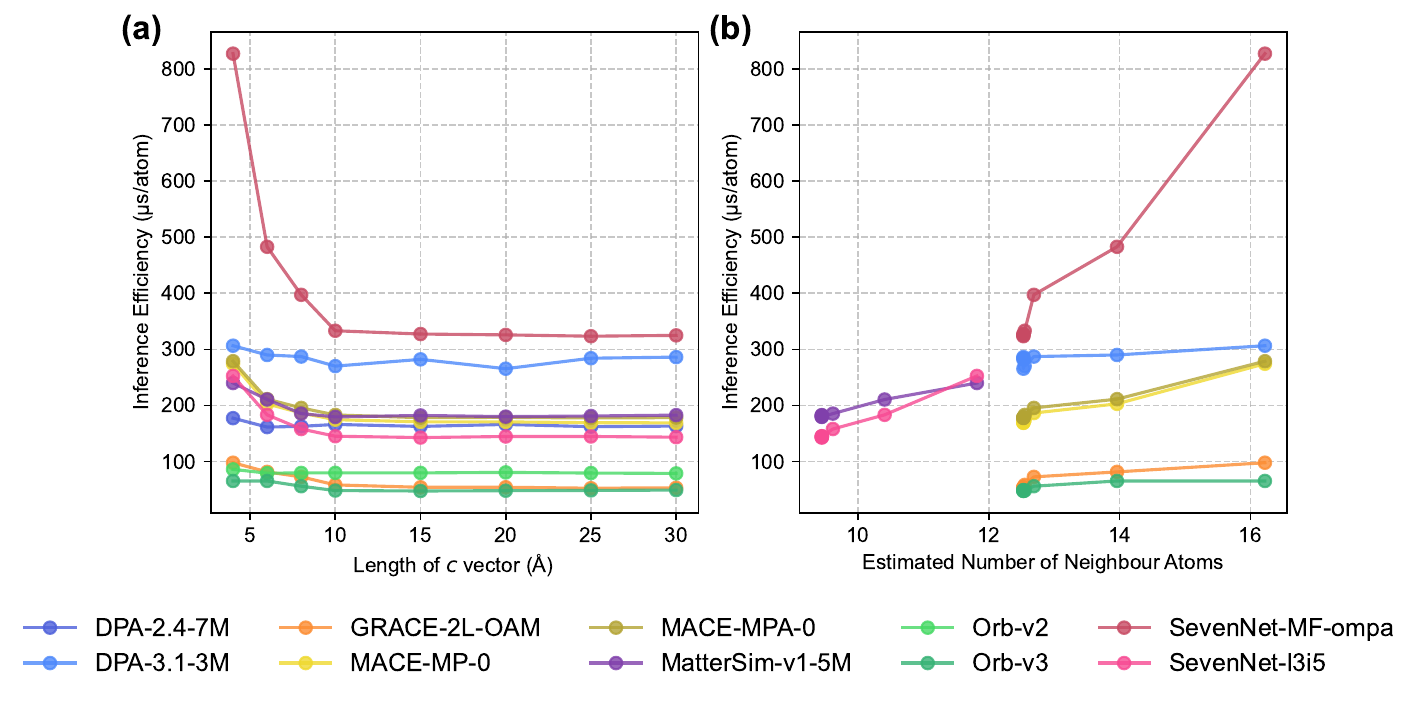}
    \caption{Effect of the number of neighbor atoms on inference time, (a) correlation between inference time and vacuum thickness in a bilayer $\text{Na}_8$ 2D structure. (b) Dependence of inference time on the number of neighbor atoms in a bilayer $\text{Na}_8$ 2D structure. The number of neighbor atoms is estimated using the cutoff radius specific to each model.}
    \label{fig:na-nnei}
\end{figure}
\clearpage

\section{Stability Test Results}
\begin{table}[h]
    \centering
    \caption{NVE molecular dynamics simulations over nine atomic systems.}
    \label{tab:si-nvemd}
    {\small
    \begin{tabular}{lccc}
        \toprule
        \textbf{Model} & 
        \textbf{\makecell{Energy Drift \\ (meV/atom/ps)}} & 
        \textbf{\makecell{Success \\Rate }} &
        \textbf{\makecell{Direct Force \\ Prediction}} \\
        \midrule
        DPA-3.1-3M & 0.008  & 0.89 & No \\
        DPA-2.4-7M & 0.018  & 1 & No \\
        MACE-MP-0 & 0.005  & 1& No \\
        MACE-MPA-0 & 0.004  & 1& No \\
        Orb-v2  & 222.8  & 1& Yes \\
        Orb-v3  & 0.005  & 1& No \\
        SevenNet-l3i5 & 0.012  & 1& No \\
        SevenNet-MF-ompa  & 0.010  & 1& No \\
        MatterSim-v1-5M  & 0.007 & 1 & No \\
        GRACE-2L-OAM  & 0.010 & 1 & No \\
        \bottomrule
    \end{tabular}
    }
\end{table}
\clearpage


\def\bibsection{\section*{\refname}} 
\bibliography{ref}
\end{document}